\documentclass[pdflatex,final,10pt]{article}
\usepackage[dvips]{epsfig}
\usepackage{graphics}
\usepackage{subfigure}
\usepackage{latexsym}
\usepackage{verbatim}
\usepackage{amsmath}
\usepackage{amssymb}
\usepackage{comment}
\usepackage{units}
\usepackage[]{hyperref}
\hypersetup{
    colorlinks=true,%
    filecolor=black,%
    linkcolor=black,%
    urlcolor=black
}
\usepackage{mathtools}

\graphicspath{{fig/}}	
\usepackage{amsthm}

\newtheorem{hypothesis}{Hypothesis}[section]
\newtheorem{proposition}{Proposition}[section]
\newtheorem{remark}{Remark}[section]
\newtheorem{definition}{Definition}[section]

\def\theequation {\arabic{section}.\arabic{equation}}
\def\eps{\varepsilon}
\def\Id{\mathbb{I}\textrm{d}}

\def\u{\mathbf{u}}
\newcommand{\IMT}{Universit\'e de Toulouse; UPS, INSA, UT1, UTM, Institut de Math\'ematiques de Toulouse,
CNRS, Institut de Math\'ematiques de Toulouse UMR 5219, F-31062 Toulouse, France; fabrice.deluzet@math.univ-toulouse.fr}
\newcommand{\HIT}{School of Mathematics, Harbin Institute of Technology, 92 West Dazhi Street, Nan Gang District, Harbin, 150001, China;
yangchang@hit.edu.cn}
\newcommand{\Dollars}{This work has been supported by the Natural Science Foundation of China (12371432) and  a public grant from the "Laboratoire d'Excel\-lence Centre International de Math\'ematiques et d'Informatique"  (Labex CIMI) overseen by the French National Agency (ANR) as part of the "Investissement d'Avenir" program (reference ANR-11-LABX-0040) in the frame of the PROMETEUS project (PRospect of nOvel nuMerical modEls for elecTric propulsion and low tEmperatUre plaSmas).\newline
  Support from the ``F\'ed\'eration de Fusion pour la Recherche par Confinement Magn\'etique'' (FrFCM) in the frame of the project ``NEMESIA: Numerical mEthods for
  Macroscopic models of magnEtized plaSmas and related anIsotropic equAtions'' is also acknowledged.\newline
  CY is supported by the Fog Research Institute under contract no. FRI-454.
  CY is also supported by  the Fundamental Research Funds for the central universties (2022FRFK060021).\newline
  FD acknowledges invitations as a scholar professor by Harbin Institute of Technology in 2023.\newline
  CY acknowledges invitations as a scholar professor by Labex CIMI during 2023 and 2024.\newline
  
  The authors would like to thank Prof. Xiaogang Wang and Prof. Tianchun Zhou for stimulating exchanges and Tao Huang and Min Wang for providing insights in the operating of the HIT-PSI linear plasma experiment.
  }

\begin{document}
\title{Multiscale numerical methods for isothermal fluid models of confined plasmas
\footnote{\Dollars}}
\author{C. Yang\thanks{\HIT}
\and F. Deluzet\thanks{\IMT}}
\newtheorem{assumption}{Assumption}

\author{Chang Yang\thanks{\HIT} \and Fabrice Deluzet\thanks{\IMT}}

\hyphenation{bounda-ry rea-so-na-ble be-ha-vior pro-per-ties
cha-rac-te-ris-tic}

\maketitle

\begin{abstract}
 The aim of this work is to introduce a numerical method to cope with the multiscale nature of confined plasma physics. These investigations are focused on fluid plasma description under large magnetic field. The difficulties in this context stem from intense magnetization of the plasma, inducing a severe anisotropy, possible quasi-neutrality breakdowns, which may occur locally in the plasma and, eventually, the drift regime which prevails for the description of the electrons. These characteristics bring small parameters compared to the scale of the studied device. This work is therefore devoted to highlighting the difficulties specific to this context and to developing numerical methods efficient to cope with this multiscale nature of the physics within the framework of asymptotic-preserving methods.
\end{abstract}

\vspace{0.1cm}

\noindent 
{\small\sc Keywords.}  {\small Drift approximation; Quasi-neutrality; Anisotropy; Asymptotic Preserving schemes}

\tableofcontents

\section{Introduction}
\label{sec:introduction}
The aim of this work is to develop robust simulation tools for investigating the physics of the linear plasma device HIT-PSI \cite{wang2023}, which has been designed to study fusion plasmas and reduce heat flow at the divertor target. In doing so, several challenges encountered by numerical methods in efficiently modeling the behavior of hot plasmas under strong magnetic fields are addressed.

Magnetized plasma models involve multiple scales  \cite{chodura_plasma_1986} with varying magnitudes  during plasma evolution. In this context, asymptotic-preserving schemes \cite{jin_efficient_1999,jin_asymptotic_2012,degond_asymptotic-preserving_2013,degond_asymptotic-preserving_2017} have been shown to be highly effective for managing both the multiscale nature of the problem and the transitions between different regimes. One of the most common regime transitions in fluid plasma models is related to plasma quasi-neutrality—which can break down locally in both time and space \cite{degond_plasma_2003,degond_numerical_2012,degond_numerical_2012-1,crispel_asymptotic_2007,degond_analysis_2008,alvarez-laguna_plasma-sheath_2019,crestetto_bridging_2020}—and to the drift approximation \cite{badsi_study_2019,degond_asymptotic_2009,brull_asymptotic-preserving_2011}, also known as the low Mach regime in the context of fluid dynamics \cite{degond_mach-number_2007,herbin_staggered_2012,tang_second_2012}. 

The objective herein is to introduce an efficient numerical method for simulating plasma dynamics under strong magnetic fields, capable of managing local quasi-neutrality breakdowns and moving beyond the drift regime. This work contributes therefore to the development of Asymptotic-Preserving methods capable of handling two distinct singular limits. Previous instances include the quasi-neutral and fluid limits \cite{crouseilles_multiscale_2016}, the fluid and high-field limits \cite{crouseilles_asymptotic_2011}, and, more closely related to the present study, the quasi-neutral and low-Mach limits \cite{alvarez-laguna_asymptotic_2020}. This list is non-exhaustive.

A key contribution of this analysis is the consideration of errors arising from computer arithmetic and the numerical solution of linear systems resulting from various discretizations, in order to assess the asymptotic-preserving property of the proposed schemes. Particular attention is given to the amplification of these errors by the asymptotic parameter that defines the drift limit. Consequently, discretization parameters must be carefully chosen to mitigate this amplification and preserve numerical accuracy.
This issue has received little attention in the literature, where most studies have focused on the derivation, analysis, and numerical investigation of one-dimensional problems—cases in which direct solvers typically produce residual errors close to machine precision, effectively masking such error amplifications. In contrast, large-scale three-dimensional computations often rely on iterative solvers, which rarely achieve such precision. This highlights the need to develop discretizations that are robust against numerical error amplification, including errors introduced by the solution of linear systems.
Furthermore, this analysis reveals that the discretization of second-order differential operators, arising from implicit schemes within colocated spatial discretizations, can break the asymptotic-preserving property of a method. This underscores the fact that the asymptotic-preserving nature of a scheme cannot be determined solely by its time discretization.

The paper is organized as follows. Sec.~\ref{sec:Modelling} introduces a simplified model designed to highlight the challenges associated with simulating hot plasma confined by a strong magnetic field. The characteristic scales are set according to the HIT-PSI experiment, providing a reference for the magnitude of various parameters and leading to a dimensionless formulation of the governing equations used for discretization analysis.
Sec.~\ref{sec:num:meth} details the time and space discretization schemes, with a particular emphasis on how discrete quantities preserve asymptotic equilibria. A numerical method is introduced that ensures the asymptotic-preserving property not only with respect to classical time and space discretization but also concerning the residuals arising from the numerical solution of the linear system associated with implicit discretizations.
The effectiveness of this asymptotic-preserving method is evaluated in Sec.~\ref{sec:results} through various numerical investigations, including computations in the drift regime and cases involving local quasi-neutrality breakdowns. Finally, conclusions and future research directions are presented in Sec.~\ref{Conclusions}.

\section{Modelling}
\label{sec:Modelling}
\subsection{An isothermal electrostatic model}
The starting point is a minimal plasma description. An isothermal fluid model is therefore considered for both electrons and the ions with an electrostatic approximation for the electromagnetic field. Let $n_i, n_e$ be the ion and electron densities, $\mathbf{u}_i, \mathbf{u}_e$ the particles' velocities, $P_i,T_i,m_i$ $P_e,T_e,m_e$ the pressure, temperature and mass of the species, $E$ and $B$ denote the electric and magnetic fields. Note, that, owing to the isothermal assumption of the model, the temperatures $T_i, T_e$ are constant. The exterior and constant magnetic field is also assumed strong enough for the field induced by the motion of particles to be neglected hence the electrostatic assumption. Additionally, the elementary charge $e$ is introduced along with $\varepsilon_0$, the vacuum permittivity, $k_B$, the Boltzmann constant, to write the bi-fluid isothermal electrostatic model:
\begin{subequations}\label{eq:isothermal}
  \begin{align}
    &-\Delta\phi = \frac{e}{\varepsilon_0} (n_i - n_e),
    \end{align}
\begin{align}
&\frac{\partial}{\partial t}n_\alpha + \nabla\cdot(n_\alpha \mathbf{u}_\alpha)=S_\alpha,\\
&\begin{multlined}[0.9\textwidth]
  m_\alpha\left[ \frac{\partial}{\partial t}(n_\alpha\mathbf{u}_\alpha) + \nabla\cdot(n_\alpha \mathbf{u}_\alpha   \otimes\mathbf{u}_\alpha) \right] + \nabla P_\alpha=\\
    q_\alpha n_\alpha (E + \mathbf{u}_\alpha\times B)- m_\alpha \nu_\alpha n_\alpha \mathbf{u}_\alpha, 
\end{multlined}
\end{align}
%
where $\alpha$ denotes the particle specie, $\alpha=e$ for the electrons and $\alpha=i$ for the ions,
$ P_\alpha = n_\alpha k_B T_\alpha$ is the pressure of the specie $\alpha$ and $q_\alpha$ is the particle charge, $q_e = -e $ and $q_i =e$. 
\end{subequations}

The neutral particles' influence on the plasma evolution is modelled by means of the source terms in both the density and the momentum conservation equations: $S_\alpha$ is the rate of plasma density created by ionization or destroyed by recombination, $\nu_\alpha$ being the collision frequency of the specie $\alpha$ against neutral particles assumed to be at rest.

To highlight the multiscale nature of the model~\eqref{eq:isothermal}, the equations are rescaled thanks to the  characteristic quantities listed in Tab.~\ref{tab:Characteristic_quantities} with the following definition for the Debye length $\lambda_D$, the cyclotron frequency $\omega_{c,\alpha}$ and the speed of sound $C_{s,\alpha}$:
\begin{equation}
  \lambda_D = \left( \frac{\varepsilon_0 k_B \bar{T}_e}{e^2\bar{n}} \right)^{1/2} \,, \qquad \omega_{c,\alpha} =\frac{e\bar{B}}{m_\alpha} \,, \qquad   C_{s,\alpha} = \left(\frac{k_B \bar{T}_\alpha}{m_\alpha} \right)^{1/2}\,.
\end{equation}
\begin{table}[htbp]
  \begin{center}
\caption{Characteristic quantities and dimensionless parameters.\label{tab:Characteristic_quantities}}
\begin{tabular}{|ll|}
\hline
\multicolumn{1}{|c|}{Characteristic quantity} & \multicolumn{1}{|c|}{Definition} \\
\hline\hline
$\bar{x}$ & Characteristic length;\\
$\bar{t}$ & Characteristic time;\\
$\bar{u}={\bar x}/{\bar t}$ & Characteristic velocity;\\
$\bar{n}$ & Characteristic particle density;\\
$ \bar{E}={\bar{\phi}}/{\bar{x}}$ & Electric field scale;\\
$\bar{B}$ & External magnetic field;\\
$\bar{S}={\bar n}/{\bar t}$ & Density creation/depletion rate;  \\
$ \bar{T}_\alpha$ &  Typical temperature (species $\alpha$);\\
$C_{s,\alpha}$ & Speed of sound (species $\alpha$). \\ \hline \hline

\multicolumn{1}{|c|}{Dimensionless Parameter} & \multicolumn{1}{|c|}{Definition} \\ \hline \hline
$ \lambda=\lambda_D/\bar{x}$ & Rescaled Debye length;\\
$ \eta=(e\bar{\phi})/(k_B \bar{T}_e)$ & Electric to electron internal energies ratio;\\
$ M_\alpha={\bar u}/{C_{s,\alpha}}$ & Mach number (species $\alpha$);\\
$ \kappa_\alpha={\bar \nu_\alpha}/ \bar{t}$ & Number of collisions during a typical time;\\
$\Omega_\alpha=\omega_{c,\alpha}\bar{t}$ & Number of cyclotron rotations per typical time.\\
\hline
\end{tabular}
\end{center}
\end{table}

For conciseness of the notations, the same symbols are conserved for the dimensionless variables, yielding the following rescaled model:%
\begin{subequations}\label{dimensionlessEP} 
\begin{align}
&\displaystyle \frac{\partial}{\partial t}n_i + \nabla\cdot(n_i \mathbf{u}_i)=S_i,\\
&\begin{aligned}
M_i^2\left[ \frac{\partial}{\partial t}(n_i\mathbf{u}_i) + \nabla\cdot(n_i \mathbf{u}_i   \otimes\mathbf{u}_i) \right] +& \nabla (n_i T_i)=\\& \eta n_i E + M_i^2\Omega_i n_i\mathbf{u}_i\times B-M_i^2\kappa_i n_i\mathbf{u}_i,\\
\end{aligned}\\
%
&\displaystyle \frac{\partial}{\partial t}n_e + \nabla\cdot(n_e \mathbf{u}_e)=S_e,\\
&\begin{aligned}
M_e^2\left[ \frac{\partial}{\partial t}(n_e\mathbf{u}_e) + \nabla\cdot(n_e \mathbf{u}_e   \otimes\mathbf{u}_e) \right] + &\nabla (n_e T_e)=\\ -&\eta n_e E - M_e^2 \Omega_e n_e \mathbf{u}_e\times B-M_e^2\kappa_en_e\mathbf{u}_e,
\end{aligned}
\end{align}
\begin{equation}
  -\lambda^2\eta\Delta\phi =n_i - n_e\,.\label{eq:Poisson}
\end{equation}
\end{subequations}
\subsection{Scaling relations and asymptotic regimes}
The applications targeted are related to the linear plasma device facility HIT-PSI~\cite{wang2023} operated to study the parallel dynamic of magnetically confined plasma. In this experiment, collisions with neutral particles are spatially localized and therefore do not uniformly impact plasma evolution. As a result, they are omitted in the initial analysis. The main characteristics, gathered in Appendix~\ref{sec:RQ} (see in particular Tab~\ref{tab:physics}), yield the following scaling relations:
\begin{itemize}
  \item The electrons are mainly in a low Mach regime with $M_e^2\sim 10^{-6}$ while the inertia of the ions may not necessarily be neglected $M_i^2\sim 10^{-2}$.  This electron scale will be attached to the parameter $\eps_{M,e}=M_{e}^2\ll 1$.
  Letting $\eps_{M,e}\to 0$, assuming the ratio $\eps_{\Omega,e}/\eps_{M,e}$ remains finite, the so-called diamagnetic and $E\times B$ drifts are recovered for the perpendicular component of the electron velocity
  \begin{subequations}\label{eq:drift:balance}
  \begin{equation}\label{eq:perpendicular:drifts}
    u_{e,\perp} =  \left[\frac{\eps_{\Omega,e}}{\eps_{M,e}}\right] \left(\frac{\nabla_\perp P_e\times B}{n_e |B|^2} + \eta \frac{E\times B}{|B|^2}\right) \,.
  \end{equation}
  Along the magnetic field lines, the Botlzmann equilibrium is imposed
  \begin{equation}
    \frac{1}{n_e}\frac{\partial }{\partial y} (n_e) = \frac{\eta}{T_e} \, {\partial_y \phi} \,.
  \end{equation}
\end{subequations}
From this relation the value of the parameter $\eta$ may be inferred. Indeed, $\eta \gg 1$ entails $\partial_y \phi = 0$ hence the magnetic field lines are equipotential for the electric field, in other words, the electric field is perpendicular to the magnetic field. In the converse asymptotic $\eta \ll 1$, no parallel pressure gradients are possible. The richest parallel equilibrium is the one defined by intermediate values of this parameter $\eta \sim 1$ permitting the electric field to adjust to pressure imbalances along the magnetic field lines. This is the choice retained throughout the rest of this work.
\item The plasma is strongly magnetized owing to the smallness of the dimensionless cyclotron frequencies $\Omega_e=10^{-8}$ and $\Omega_i=10^{-5}$. The asymptotic parameters $\eps_{\Omega,e}$ and $\eps_{\Omega_,i}$ are therefore introduced, the limit $\eps_{\Omega,\alpha}=\Omega_{\alpha}^{-1} \to 0$ relating an increase in the magnetic field magnitude. The vanishing of the perpendicular velocity outlines that the typical scales are chosen accordingly to the parallel dynamics much faster than the perpendicular one.
Letting $\eps_{\Omega,e}\to 0$, Eq.~\eqref{eq:perpendicular:drifts} yields $u_{e,\perp} \to 0$ which translates zero displacements across the magnetic field lines hence a perfect confinement of the plasma.
\item Different plasma areas are identified: a dense plasma region defining a particle beam characterized by a scaled Debye length $\lambda$ as small as $10^{-6}$. The limit $\lambda\to 0$ entails the quasi-neutrality of the plasma, by means of the degeneracy of the Poisson equation~\eqref{eq:Poisson} yielding $n_e = n_i$. Nonetheless, the quasi-neutrality does not prevail uniformly in the entire device, the vessel containing the beam is indeed mainly filled with a quasi-vacuum area. Local quasi-neutrality breakdowns shall therefore be imbedded into the model.
\end{itemize}

\section{Multiscale numerical methods}\label{sec:num:meth}
\subsection{Time discretization}
The aim now is to design a numerical method encompassing the drift approximation together with local quasi-neutrality breakdowns under the severe anisotropy induced by the magnetic field. The difficulty stemming from the vanishing of parameters and the degeneracy of some equations may be addressed through the design of asymptotic-preserving schemes. Such schemes may be obtained thanks to a reformulation of the equations which turns out to an implicit discretization of carefully chosen terms. It is therefore quite natural to investigate the derivation of the multiscale numerical methods thanks to a time semi discrete system. 
To simplify this analysis, the ions are assumed to be at rest, the only electron system is therefore considered by means of the simplified set of equations
\begin{subequations}\label{eq:syst:discret}
 \begin{align}
 &-\lambda^2\Delta\phi^{k+1} =n_i^{k+1} - n_e^{k+1}\,,\label{eq:discret_poisson}\\
&\quad n_e^{k+1} = n_e^k - \Delta t \nabla\cdot(n_e \mathbf{u}_e)^{k+1}\label{eq:discret_ne}\,,\\
 &\quad  \begin{multlined}[0.8\textwidth]
    (n_e\mathbf{u}_e)^{k+1} = (n_e\mathbf{u}_e)^{k} - \Delta t\nabla\cdot(n_e \mathbf{u}_e   \otimes\mathbf{u}_e)^k \\[\jot]
    - \displaystyle \frac{\Delta t}{\varepsilon_{M,e}}\left( \nabla P_e^{k+1} + n_e^k E^{k+1}\right) - \frac{\Delta t}{\varepsilon_{\Omega,e}} (n_e\mathbf{u}_e)^{k+1}\times B.\label{eq:discret_nue} 
  \end{multlined}
\end{align}
\end{subequations}
This definition of the semi-discrete system calls few comments. 
\newline First, to recover the drifts for the perpendicular electron velocity as stated by Eq.~\eqref{eq:perpendicular:drifts}, an implicit momentum is mandatory in the definition of the Lorentz force within Eq.~\eqref{eq:discret_nue}.
\newline Second, another level of implicitness is mandatory to cope with the quasi-neutral limit. Indeed, letting $\lambda \to 0$ in Eq.~\eqref{eq:discret_poisson} entails the quasi-neutrality condition $n_e^{k+1}=n_i^{k+1}$ but, the potential cannot be explicitly obtained from this degenerate equation. Therefore, another means of computing $\phi^{k+1}$ is mandatory to cope with this regime. This is classically achieved, in the framework of Asymptotic-Preserving methods (see for instance \cite{degond_asymptotic_2009,degond_asymptotic-preserving_2017}), by substituting Poisson's equation by an equivalent reformulated one, degenerating, in the quasi-neutral limit, into $\nabla\cdot J=0$, $J$ being the particle current. The electric potential is deduced from this divergence free current constraint when $\lambda$ is set to 0. The reformulated equation is obtained thanks to an implicit discretization of both the density flux and the electric force within respectively the density conservation equation~\eqref{eq:discret_ne} and the momentum equation~\eqref{eq:discret_nue}. 
\newline Third, since the electrons are in a drift regime, the pressure shall be implicit in the momentum equation~\eqref{eq:discret_ne} to obtain stability of the discrete model irrespective to $\eps_{M,e}$-values. This issue will be detailed in the sequel.

Introducing $ \mathbb{M}_B $ the matrix satisfying $\mathbb{M}_B v = v \times B$ for any $v\in\mathbb{R}^3$ and assuming $B=(0,1,0)^T$ the momentum equation~\eqref{eq:discret_nue} is recast into
\begin{subequations}\label{sys:discret_nue2}
\begin{equation}\label{eq:discret_nue2}
  \begin{aligned}
(n_e\mathbf{u}_e)^{k+1} &= \mathbb{A}_e \left[  (\widetilde{n_e\mathbf{u}_e})^{k} - \frac{\Delta t}{\varepsilon_{M,e}} \left(\nabla P_e^{k+1} +  n_e^k E^{k+1}\right) \right],\\
(\widetilde{n_e\mathbf{u}_e})^{k}&={(n_e\mathbf{u}_e)^{k}} - \Delta t\nabla\cdot(n_e \mathbf{u}_e   \otimes\mathbf{u}_e)^k\,,
  \end{aligned}
\end{equation}
with
\begin{align}\SwapAboveDisplaySkip
  \mathbb{A}_e =
  \left( Id + \frac{\Delta t}{\varepsilon_{\Omega,e}}\mathbb{M}_B\right)^{-1}=
  \begin{pmatrix}
    \displaystyle \frac{\varepsilon_{\Omega,e}^2}{\varepsilon_{\Omega,e}^2+\Delta t^2} & 0 & \displaystyle\frac{\Delta t \,\varepsilon_{\Omega,e}}{\varepsilon_{\Omega,e}^2+\Delta t^2}\\
    0 & 1 & 0\\
    \displaystyle-\frac{\Delta t \,\varepsilon_{\Omega,e}}{\varepsilon_{\Omega,e}^2+\Delta t^2} & 0 & \displaystyle\frac{\varepsilon_{\Omega,e}^2}{\varepsilon_{\Omega,e}^2+\Delta t^2}
    \end{pmatrix}.
\end{align}
\end{subequations}
This yields the following definition of the implicit density
\begin{subequations}\label{sys:schema:imp}
 \begin{eqnarray}
n_e^{k+1} = n_e^k - \Delta t \nabla\cdot \left(\mathbb{A}_e \left[  (\widetilde{n_e\mathbf{u}_e})^{k} - \frac{\Delta t}{\varepsilon_{M,e}} \left( \nabla P_e^{k+1} + n_e^k E^{k+1}\right) \right]\right).\label{eq:discret_ne3} 
\end{eqnarray}
Substituting Eq.~\eqref{eq:discret_ne3} into Eq~\eqref{eq:discret_poisson} provides the so-called "reformulated Poisson" equation \cite{degond_asymptotic-preserving_2013,degond_analysis_2008}, classically implemented in Asymptotic-Preserving schemes to carry out the electric potential:
\begin{equation}
\begin{multlined}[0.85\textwidth]
-\nabla\cdot\left(\left[  \lambda^2 \Id 
+ \frac{\Delta t^2}{\varepsilon_{M,e}}n_e^k\mathbb{A}_e \right]\nabla\phi^{k+1}\right)  = n_i^{k+1} - n_e^k
\\
 + \Delta t \nabla\cdot \left(\mathbb{A}_e \left[ (\widetilde{n_e\mathbf{u}_e})^{k} - \frac{\Delta t}{\varepsilon_{M,e}} \nabla P_e^{k+1}  \right]\right).
\end{multlined}\label{eq:discret_refpoisson}
\end{equation}
In the limit $\lambda\to0$ the reformulated equations is not degenerate. It provides a means of computing the electric potential to enforce the balance $n_i^{k+1}-n_e^{k+1}=0$, hence the quasi-neutrality of the plasma.
\end{subequations}

The issue related to the low Mach regime may now be examined. Projecting the electron momentum conservation equation onto the parallel direction and the perpendicular ones, the following system is recovered:
\begin{subequations}\label{sys:momentum}
  \begin{align}
    (n_e\mathbf{u}_e)^{k+1}_\parallel &=  (\widetilde{n_e\mathbf{u}_e})^{k}_\parallel - \frac{\Delta t}{\varepsilon_{M,e}} \left(\nabla_\parallel P_e^{k+1} +  n_e^k \nabla_\parallel \phi^{k+1}\right)\,,\label{sys:momentum:parallel}\\
    (n_e\mathbf{u}_e)^{k+1}_\perp &= \mathbb{A}_e^\perp \left[  (\widetilde{n_e\mathbf{u}_e})^{k}_\perp - \frac{\Delta t}{\varepsilon_{M,e}} \left(\nabla_\perp P_e^{k+1} +  n_e^k \nabla_\perp \phi^{k+1}\right) \right]\,,
  \end{align}
  where,
  \begin{equation}
    \begin{aligned}
      (n_e\mathbf{u}_e)_\parallel &= (n_e\mathbf{u}_e)_y \,,& \qquad \nabla_\parallel \phi&= \partial_y \phi \,,\\
      (n_e\mathbf{u}_e)_\perp &= \big((n_e\mathbf{u}_e)_x,(n_e\mathbf{u}_e)_z\big)^T\,,& \qquad \nabla_\perp \phi &= (\partial_x\phi, \partial_z\phi)^T \,,
    \end{aligned}
  \end{equation}
and
\begin{equation}\label{eq:perpendicular:mobility:matrix}
  \mathbb{A}_{e}^{\perp} =
  \begin{pmatrix}
    \displaystyle \frac{\varepsilon_{\Omega,e}^2}{\varepsilon_{\Omega,e}^2+\Delta t^2} & \displaystyle\frac{\Delta t \,\varepsilon_{\Omega,e}}{\varepsilon_{\Omega,e}^2+\Delta t^2}\\
    \displaystyle-\frac{\Delta t \, \varepsilon_{\Omega,e}}{\varepsilon_{\Omega,e}^2+\Delta t^2} & \displaystyle\frac{\varepsilon_{\Omega,e}^2}{\varepsilon_{\Omega,e}^2+\Delta t^2}
    \end{pmatrix}.
\end{equation}
\end{subequations}
The density conservation stated by Eq.~\eqref{eq:discret_ne3} may be recast into 
\begin{subequations}\label{sys:discret_ne:accoustic}
\begin{equation}\label{eq:discret_ne:accoustic}
  \begin{multlined}[0.85\textwidth]
    n_e^{k+1} -  \frac{\Delta t^2}{\varepsilon_{M,e}} \nabla_\parallel \cdot\left( T_e \nabla_\parallel n_e^{k+1}\right) = n_e^k - \Delta t \nabla_\parallel \cdot \left( (\widetilde{n_e\mathbf{u}_e})^{k}_\parallel - \frac{\Delta t}{\varepsilon_{M,e}}  n_e^k E^{k+1}_\parallel\right) \\ 
    - \Delta t\nabla_\perp\cdot \left(\mathbb{A}_{e}^{\perp} \left[  (\widetilde{n_e\mathbf{u}_e})^{k}_\perp - \frac{\Delta t}{\varepsilon_{M,e}} \left( \nabla_\perp P_e^{k+1} + n_e^k E^{k+1}_\perp\right) \right]\right).
  \end{multlined}
\end{equation}
\end{subequations}
The left-hand side of this equation is an implicit discretization of a wave equation. Substituting an explicit discretization of the pressure in Eq.~\eqref{eq:discret_nue} to the proposed implicit scheme, would give rise to an explicit discretization of the left-hand side of Eq.~\eqref{eq:discret_ne:accoustic}. The stability of an explicit scheme is subjected to the condition $\mathcal{C} \Delta t^2 < \eps_{M,e} \Delta y^2 $, $\Delta y$ denoting the typical mesh size and $\mathcal{C}$ a constant, introducing a stringent constraint on the time step in the drift regime $\eps_{M,e} \ll 1$. Note that, restoring the dimensional quantities, this constraint writes $\mathcal{C} C_{S,e}\Delta t <\Delta y$: the drift approximation is closely related to the propagation of acoustic waves at a speed much larger than the transport one ($\sim \Delta t/ \Delta y$). The implicit discretization of the pressure gradient permits to avoid such a stability constraint by means of an implicit resolution of Eq.~\eqref{eq:discret_ne3}.

The asymptotic properties of the scheme may be outlined thanks to the investigation of a simplified framework consisting of a one dimensional discrete system reduced to the aligned direction. To this aim, second order (space) differential operator are introduced.
\begin{definition}[Elliptic operators] For any smooth function $f$, the following operators are defined
  \begin{equation}\label{eq:def:elliptic:operator}
    \begin{gathered}
      - \Delta_{\parallel,n_e^k} f := - \nabla_\parallel \cdot \left(n_e^k \nabla_\parallel f \right) \,,\qquad  - \Delta_{\parallel,T_e} f := - \nabla_\parallel \cdot \left(T_e \nabla_\parallel f \right) \,, \\
      - \Delta_{\parallel} f :=- \nabla_\parallel \cdot \left(n_e^k \nabla_\parallel f \right) \,.
    \end{gathered}
  \end{equation}
\end{definition}
These operators are assumed to define well posed problems as specified by the following hypothesis.
\begin{hypothesis}\label{hyp:ellipticity:parallel}
  The operator $-\Delta_\parallel$,  $-\Delta_{\parallel,n_e^k}$ and $-\Delta_{\parallel,T_e}$ are assumed to be elliptic.
  \end{hypothesis}
  These assumptions amount to supplement the second order operators defined by Eq.~\eqref{eq:def:elliptic:operator} with convenient boundary conditions and ensure that each implicit equation yields a well posed problem. The purpose here is not to elaborate the optimal assumptions, rather unravel the properties of the different discretizations within the same framework, as outlined by the following result.

\begin{definition}[Implicit Scheme] The implicit scheme is defined  by the system of equations composed of the reformulated Poisson equation \eqref{eq:discret_refpoisson}, coupled to the implicit density equation \eqref{eq:discret_ne:accoustic}. The momentum is recovered thanks to Eqs.~\eqref{sys:momentum}.
\end{definition}

 \begin{proposition}\label{prop:implicit:basic:singular}
  Under the ellipticity assumptions of Hyp.~\ref{hyp:ellipticity:parallel}, the implicit scheme defined 
   by Eqs.~(\ref{eq:discret_refpoisson}, \ref{eq:discret_ne:accoustic}) is well posed in the quasi-neutral limit $\lambda \to 0$ but singular in the drift limit $\eps_{M,e}\to 0$.
\end{proposition}
\begin{proof}[Proof of Prop.~\ref{prop:implicit:basic:singular}] Setting  $\xi = \varepsilon_{M,e}/\Delta t^2$, the implicit scheme may be recast into
    \begin{equation}\label{eq:schema:implicit:matrix:a}
      \begin{pmatrix}
       \displaystyle  -\xi\lambda^2 \Delta_{\parallel} -\Delta_{\parallel,n_e^k} & \displaystyle - \Delta_{\parallel,T_e} \\[0.5em]
       \displaystyle- \Delta_{\parallel,n_e^k} & \displaystyle \xi\Id -   \Delta_{\parallel,T_e} 
      \end{pmatrix}\begin{pmatrix}
        \phi^{k+1}\\[.5em] n_{e}^{k+1}
      \end{pmatrix}= \xi\begin{pmatrix}
        n_{i}^{k+1}-\widetilde{n}_{e}^{k}\\[.5em]
        \widetilde{n}_{e}^{k}
      \end{pmatrix}=\xi\begin{pmatrix}
        F_1\\[.5em]
        F_2
      \end{pmatrix}\,.
    \end{equation}
This linear system is equivalent to solving the following one
\begin{equation*}
  \begin{pmatrix}
    \displaystyle - \lambda^2 \Delta_\parallel  & 
    \displaystyle - \Id \\[0.3em]
     \displaystyle -\Delta_{\parallel,n_e^k} & \displaystyle \xi \Id -  \Delta_{\parallel,T_e}
   \end{pmatrix} \begin{pmatrix}
     \phi^{k+1} -n_{e}^{k+1}\\[0.4em]
     n_{e}^{k+1}
   \end{pmatrix} =\begin{pmatrix}
       F_1-F_2 \\
       \xi F_2
   \end{pmatrix}\,.
\end{equation*}
Simple algebra yields
\begin{equation*}
  \left\{\begin{aligned}
   \Big( -\Delta_{\parallel,n_e^k} - \lambda^2\left(\xi \Id - \Delta_{\parallel,T_e}\right) \Delta_\parallel \Big) \left(\phi^{k+1} - n_e^{k+1}
   \right) &= \xi F_1 - \Delta_{\parallel,T_e} \left( F_1 -F_2 \right) \,, \\
    \left(\xi \Id - \Delta_{\parallel,T_e}\right) n_{e}^{k+1} &= \xi F_2 + \Delta_{\parallel,n_e^k} \phi^{k+1} \,.
  \end{aligned}\right.
\end{equation*}
The first equation defines a well posed problem for $(\phi^{k+1}-n_e^{k+1})$ for $\lambda=0$.  For $\lambda >0 $ this result will be assumed. The
density $n_{e}^{k+1}$ is carried out by means of the second equation.

Letting now $\eps_{M_e}\to 0$ or equivalently $\xi \to 0$ in Eq.~\eqref{eq:schema:implicit:matrix:a} the system yields  a matrix with both lines linearly dependent, hence the singularity of the scheme in the drift limit.
\end{proof}
The singularity of the implicit scheme is explained by the use of the momentum equation in the reformulated Poisson equation. The degeneracy of this equation in the drift limit leads to the singularity of the problem. A workaround consists in not substituting the implicit density within Poisson's equation, thereby discarding the use of the momentum equation to derive this reformulated equation.
\begin{definition}[Alternative Implicit Scheme] An alternative implicit scheme is defined by the following set of equations
  \begin{subequations}\label{eqs:alternative:implicit:scheme}
    \begin{align}
      n_e^{k+1} &= \lambda^2 \Delta_{\parallel} \phi^{k+1} +n_i^{k+1}\,,\label{eqs:alternative:implicit:scheme:a} \\
      -\Delta_{\parallel,n_e^k}^h \phi^{k+1}  &= - \left[\xi \Id - \Delta_{\parallel,T_e}^h\right] n_{e}^{k+1} + \xi F_2 \,.\label{eqs:alternative:implicit:scheme:b}
    \end{align}
  \end{subequations}
\end{definition}

The alternative implicit scheme is not prone to any degeneracy in either the quasi-neutral or the drift limit.
\begin{proposition}\label{prop:alternative:implicit:basic:singular}
  Under the ellipticity assumptions of Hyp.~\ref{hyp:ellipticity:parallel}, the alternative implicit scheme defined 
   by Eqs.~\eqref{eqs:alternative:implicit:scheme} is well posed in both the quasi-neutral and drift limits $(\lambda,\eps_{M,e}) \to 0$.
\end{proposition}
\begin{proof}[Proof of Prop.~\ref{prop:alternative:implicit:basic:singular}] Setting $\lambda=0$, the electron density is carried out thanks to Eq.~\eqref{eqs:alternative:implicit:scheme:a} to match the ion density. The electric potential is the solution to Eq.~\eqref{eqs:alternative:implicit:scheme:b}. Setting $\varepsilon_{M,e}=0$ (or equivalently $\xi=0$) only changes marginally Eq.~\eqref{eqs:alternative:implicit:scheme:b}, which, owing the ellipticity assumptions, remains a well posed problem for the electric potential.
\end{proof}
\begin{remark}[A comparison with the scheme of Alvarez {\it et al} \cite{alvarez-laguna_asymptotic_2020}]\label{remark:Alvarez}
  The implicitness of the pressure term may not be mandatory to gain the stability of the scheme in the drift limit ($\varepsilon_{M,e}\to 0$) as demonstrated by the investigations conducted in \cite{alvarez-laguna_asymptotic_2020}. The discretization proposed by the authors consists in first, recasting of the electron system under a non-conservative form; second, when computing the divergence of the velocity, substituting the Laplacian of the electric potential by the source term of Poisson's equation. Using similar notations to Eqs.~\eqref{sys:discret_nue2}, this yields the following model equation:
  \begin{equation*}
    \nabla_\parallel \cdot (\mathbf{u}_e)^{k+1}_\parallel =  \nabla_\parallel \cdot (\widetilde{\mathbf{u}_e})^{k}_\parallel - \frac{\Delta t}{\varepsilon_{M,e}} \nabla_\parallel \cdot \left(\frac{T_e}{n_e^k}\nabla_\parallel n_e^{k} - \lambda^{-2}(n_e^{k+1}-n_i^k)\right)\,.
  \end{equation*}
  This relation is inserted into Eq.~\eqref{eq:discret_ne} to derive the one driving the evolution of the density. This stability of this scheme relies on the term proportional to $(\lambda^2\varepsilon_{M,e})^{-1}$, comparable to a relaxation term, stabilizing the explicit pressure. This formulation also offers the advantage of decoupling the computation of the density from that of the electric field.

  For magnetized plasmas, the electron velocity is related to the electric field through the mobility matrix as stated by Eqs.~\eqref{sys:discret_nue2}, therefore the electric force within the divergence of the velocity does not reduce to the Laplacian of the electrostatic potential. The decoupling of the equations providing the density and the electric field cannot be preserved for magnetized frameworks. This motivates the derivation of different discretizations.
\end{remark}
\subsection{Space discretization and consistency with asymptotic equilibria}
 The discretization is applied to both the time and the space variables for two dimensional problems. The discrete system, attached to the parallel direction, reduces to 
\begin{subequations}\label{sys:parallel:impilicit}
 \begin{align}
  &\begin{multlined}[0.85\textwidth]
   -\lambda^2 \Delta_\parallel^h \phi^{k+1}_h =  n_{i,h}^{k+1} - n_{e,h}^{k+1} 
    \label{sys:parallel:impilicit:a}\,,
     \end{multlined}\\
    &\begin{multlined}[0.85\textwidth]
      n_{e,h}^{k+1} -  \frac{\Delta t^2}{\varepsilon_{M,e}} \Delta_{\parallel,T_e}^h n_{e,h}^{k+1} = \widetilde{n}_{e,h}^k + \Delta t \nabla_\parallel^h \cdot \left(  \frac{\Delta t}{\varepsilon_{M,e}}  n_{e,h}^k \nabla_\parallel^h\phi^{k+1}_h\right)\,,\label{sys:parallel:impilicit:b}
    \end{multlined}\\  
    &(n_{e,h} \u_{e,h})^{k+1}_\parallel = (\widetilde{n_{e,h} \u_{e,h}})^{k}_\parallel - \frac{\Delta t}{\varepsilon_{M,e}} \left(T_e \nabla_\parallel^h n_{e,h}^{k+1} - n_{e,h}^k \nabla_\parallel^h \phi^{k+1}_h \right) \,,\label{sys:parallel:impilicit:c}
  \end{align}
  with
  \begin{align}
    &\begin{multlined}[0.8\textwidth]
      (\widetilde{n_{e,h} \u_{e,h}})^{k}_\parallel = ({n_{e,h} \u_{e,h}})^{k}_\parallel- \Delta t \nabla_\parallel^h \cdot \left( (\widetilde{n_{e,h}\mathbf{u}_{e,h}\mathbf{u}_{e,h}})^{k}_\parallel \right)\\- \Delta t\nabla_\parallel^h \cdot \left(  \beta^k_h \nabla_\parallel^h n_{e,h}^k\right)\,,\label{sys:parallel:impilicit:d}
    \end{multlined}\\
      &\widetilde{n}_{e,h}^k = n_{e,h}^k - \Delta t \nabla_\parallel^h \cdot \left( (\widetilde{n_{e,h}\mathbf{u}_{e,h}})^{k}_\parallel \right)-\Delta t \nabla_\parallel^h \cdot \Big(  \beta_h^k \nabla_\parallel^h (n_{e,h}^k\mathbf{u}_{e,h}) \Big)\,.\label{sys:parallel:impilicit:e}
  \end{align}
\end{subequations}
In these equations, the symbols denoted with a $h$ represent the numerical approximation of the continuous quantities as well as the discrete differential operators. The diffusion term in (Eqs~\ref{sys:parallel:impilicit:d}--\ref{sys:parallel:impilicit:e}), controlled by $\beta^k_h$, are artificial viscosity corrections, introduced to upwind the centered divergence operators, this feature is addressed in Def.~\ref{def:upwinding}.
\begin{definition}{(Staggered space discretization).}\label{def:staggered:operators} The unknowns are carried out on a staggered grid: the electric potential and the density,
  $\phi_h^k$ and $n_{e,h}^k$, are discretized on the grid cells, while the momentum components are carried out on the edge centers yielding
  \begin{subequations}\label{sys:def:discrete:space:opertors}
  \begin{align}
    (\nabla_\parallel^h \phi_h^k)_{i,j+1/2} &= \frac{\phi_{i,j+1}^k-\phi_{i,j}^k}{\Delta y} \,,\\
    \left(\nabla_\parallel^h \cdot (n_{e,h}^k \mathbf{u}_{e,h}^k)_\parallel\right)_{i,j} &= \frac{\big((n_{e,h}^k \mathbf{u}_{e,h}^k)_\parallel\big)_{i,j+1/2} - \big((n_{e,h}^k \mathbf{u}_{e,h}^k)_\parallel\big)_{i,j-1/2}}{\Delta y} \,;
  \end{align}
   $\psi_{i,j}$ denoting the numerical approximation of a function $\psi(x_i,y_j)$, $\zeta_{i,j+1/2}$ that of a function $\zeta(x_i,y_{j+1/2})$, $\psi_h$ is a vector of $\mathbb{R}^{N_x\times N_y}$ containing the approximations carried out on the ${N_x\times N_y}$ mesh 
  \begin{equation}
    \begin{aligned}\label{sys:def:discrete:space:opertors:d}
    &(\psi_h)_{i,j} = \psi_{i,j}\,, \qquad  &&\forall (i,j)\in [1,N_{x}]\times [1,N_{y}]\,, \\
    &(\zeta_h)_{i,j+1/2} = \zeta_{i,j+1/2}\,, \qquad  &&\forall (i,j)\in [1,N_{x}]\times [1,N_{y}-1]\,. \\
    \end{aligned}
  \end{equation}
  A similar scheme is applied in the other direction to define the entire discrete operators for three-dimensional problems.
  Eventually, the second order operators are defined thanks the discrete divergence and gradient operators:
  \begin{align}
    \Delta_{\parallel}^h \psi^h = \nabla_\parallel^h \cdot \nabla_{\parallel}^h \psi^h \,.
  \end{align}
\end{subequations}
\end{definition}
As outlined by the definition of the discrete operators, the divergence is a centered differenced approximation which shall be corrected to ensure stability. 
\begin{definition}{(Flux upwinding).}\label{def:upwinding}
  Numerical viscosity corrections are added to the divergence operator to upwind this first-order discrete operator. The approach implemented in this work uses the local Lax-Friedrichs flux, also known as the Rusanov flux \cite{rusanov_calculation_1962} or the zero-degree polynomial flux \cite{degond_polynomial_1999}. Here, the parameter $\beta_h^k$
 represents the local characteristic speed of the hyperbolic part of the system with ({\it i.e.}, the Euler equations without source terms). However, since acoustic waves are treated implicitly, the speed of sound is excluded from the definition of these characteristic speeds. This leads to the following definition:
  \begin{equation}
  \left(\beta^k_h \nabla_\parallel n_{e,h}^k\right)_{i,j+1/2} = \frac{1}{2}\left| (\mathbf{u}_{e,h}^k)_\parallel\right|_{i,j+1/2} \left((n_{e,h}^{k})_{i,j+1} - (n_{e,h}^{k})_{i,j}\right).
  \end{equation}
\end{definition}
Omitting acoustic characteristic speeds to upwind the fluxes is essential for achieving a scheme that avoids excessive diffusion in the drift regime and is commonly applied in low-Mach schemes \cite{degond_mach-number_2007,tang_second_2012,dimarco_second-order_2018}. This flux choice has been implemented in previous works \cite{degond_mach-number_2007, brull_asymptotic-preserving_2011} and could be enhanced by more sophisticated methods borrowed to low-Mach schemes (see, for instance \cite{alvarez-laguna_asymptotic_2020,badsi_study_2019,dimarco_second-order_2018} and related discussions). However, these improvements are deferred to future work.

The implicit scheme may be recast into the following linear system
\begin{equation}\label{sys:implicit:scheme}
  \begin{pmatrix}  - \lambda \Delta_{\parallel}^h & \Id \\[0.5em]
    -\xi^{-1} \Delta_{\parallel,n_e^k}^h & \Id - \xi^{-1} \Delta_{\parallel,T_e}^h
  \end{pmatrix}
 \begin{pmatrix}
  \phi^{k+1}_h\\[0.5em]
  n_{e,h}^{k+1}
 \end{pmatrix}=\begin{pmatrix}
  n_{i,h}^{k+1} \\[0.5em]
  \widetilde{n}_{e,h}^{k}
 \end{pmatrix}+\begin{pmatrix}
  R^h_\phi \\[0.5em]
  R^h_{n_e}
 \end{pmatrix}
\end{equation}
where $\xi = \varepsilon_{M,e}/\Delta t^2$ and,  $(R^h_\phi$,\,$R^h_{n_e}) \in (\mathbb{R}^{N_x\times N_y})^2$ is the residual error stemming from the linear system resolution. The introduction of this vector reflects the fact that working with the finite-precision arithmetic  of computers or numerically solving a linear system introduces an additional error on top of the discretization error.


\begin{proposition}[Discrete consistency with the quasi-neutral limit]\label{prop:quasineutral:limit}
  In the quasi-neutral limit $\lambda = 0$ the charge density, carried out by the implicit scheme defined Eq.~\eqref{sys:implicit:scheme}, is controlled by the linear system residual $R^h_\phi$
  \begin{equation}\label{eq:estimate:charge:density}
    \left\| n_{e,h}^{k+1} - n_{i,h}^{k+1} \right\|_\infty= \left\| R_\phi^h \right\|_\infty\,.
  \end{equation}  
\end{proposition}
 
\begin{proof}[Proof of Prop.~\ref{prop:quasineutral:limit}]%

  Setting $\lambda=0$ in the first line of the block system defined by Eq.~\eqref{sys:implicit:scheme} the following balance is recovered
  \begin{equation*}
    n_{e,h}^{k+1} - n_{i,h}^{k+1} = R_\phi^h \,,
  \end{equation*}
  yielding the estimate \eqref{eq:estimate:charge:density}.
\end{proof}

The consistency of the discrete system with the Boltzmann equilibrium may now be investigated. 
\begin{proposition}[Discrete consistency with the Boltzmann equilibrium]\label{prop:Bolzmann:equilibrium} Provided that the elliptic operators are issued from the discrete divergence and gradient operators as defined by Eqs.~\eqref{sys:def:discrete:space:opertors} as well as a zero flux condition, for both the density and the electric potential, is applied at one end of the magnetic field lines, the Boltzmann equilibrium issued from Eq.~\eqref{sys:parallel:impilicit:b} in the limit $\eps_{M,e} \to 0$ satisfies
  \begin{equation}\label{eq:estimate:Boltzmann:equilibrium}
   \left\|T_e \nabla_\parallel^h n_{e,h}^{k+1} - n_{e,h}^k \nabla_\parallel^h \phi^{k+1}_h \right\|_\infty \sim \left\|R_{n_e}^h\right\|_\infty\,.
  \end{equation}
\end{proposition}
\begin{proof}[Proof of Prop.~\ref{prop:Bolzmann:equilibrium}]
  Without loss of generality, the first line of the second block linear system in Eq.~\eqref{sys:implicit:scheme} is assumed to discretize the zero flux boundary condition, denoting
  \begin{equation}
    \mathcal{Q}_{i,j+1/2}^{k+1} = \left(T_e \nabla_\parallel^h n_{e,h}^{k+1} - n_{e,h}^k \nabla_\parallel^h \phi^{k+1}\right)_{i,j+1/2}
  \end{equation} 
  this boundary condition is translated into
  \begin{equation}
   \mathcal{Q}_{i,1/2} = 0 \,,
  \end{equation}
  the lines of the linear system providing
  \begin{equation}
    \frac{1}{\Delta y} \left( \mathcal{Q}_{i,j+1/2,l} - \mathcal{Q}_{i,j-1/2} \right) =\xi (\widetilde{n}_{e,h}^k-n_{e,h}^{k+1})_{i,j} +  (R_{n_e}^h)_{i,j} \,.
  \end{equation}
 Summing the $J^\text{th}$ first lines of the linear system, for $J\leq N_y$, yields
  \begin{equation}
    \mathcal{Q}_{i,J+1/2} = \xi \sum_{j=1}^{j=J} (\widetilde{n}_{e,h}^k-n_{e,h}^{k+1})_{i,j}  +  \Delta y \sum_{j=1}^{j=J} (R_{n_e}^h)_{i,j} .
  \end{equation}
The Boltzmann relation is obtained in the limit of vanishing $\xi$ (or $\varepsilon_{M,e}$), therefore the first term in the right-hand side of the preceding identity may be discarded in the limit $\xi \to 0$, yielding for  $1\leq J \leq N_y $
\begin{equation}
  | \mathcal{Q}_{i,J+1/2} |\sim \Delta y \, \Big|\sum_{j=1}^{j=J} (R_{n_e}^h)_{i,j} \Big| \leq \Delta y \sum_{j=1}^{j=J}\left|(R_{n_e}^h)_{i,j}\right|\leq N_y \Delta y \left\|R^h_{n_e}\right\|_\infty\,.
\end{equation}
\end{proof}
\begin{remark}
  A more favorable estimate may be proposed to measure the consistency of the discrete system with the Boltzmann relation. It is derived using the assumption that the residual $R_{n_e}^h$ is an unsigned quantity with all its components with the same magnitude. This is the property that may be assumed for residuals issued from the resolution of a linear system by a direct method. This entails the following estimate
\begin{equation}\label{eq:unsigned:residual:estimate}
  | (R_{n_e}^h)_{i,J} | \sim \left| \sum_{m=1}^{m=J}(R_{n_e}^h)_{i,m} \right| \,, \qquad  1\leq J \leq N_y \,.
\end{equation}
This finally provides the following estimate
  \begin{equation}\label{eq:estimate:Bontzmann:balance}
    \left|T_e \nabla_\parallel^h n_{e,h}^{k+1} - n_{e,h}^k \nabla_\parallel^h \phi^{k+1}_h \right| \sim \Delta y \left|R^h_{n_e}\right|\,.
  \end{equation} 
\end{remark}

The estimate of the Boltzman equilibrium satisfied by the discrete quantities is central in the analysis of these discretizations. Indeed, the main difficulty here is related to the computation of the parallel momentum by means of Eq.~\eqref{sys:parallel:impilicit:c} characterized in the following lines.
\begin{proposition}\label{prop:unbounded:parallel:momentum}
  Assuming that the estimate stated by Eq.~\eqref{eq:estimate:Boltzmann:equilibrium} holds true, the parallel momentum issued from Eq.~\eqref{sys:parallel:impilicit:c} is not bounded in the limit $\eps_{M,e}\to 0$,
  \begin{equation}
    \left\|\frac{1}{\Delta t}\left({(n_e\u_{e,h})^{k+1}_\parallel - (\widetilde{n_e\u_{e,h}})^{k}_\parallel}\right)\right\|_\infty \sim \frac{{1}}{\varepsilon_{M,e}} \left\|R_{n_e}^h \right\|_\infty \,.
  \end{equation}
\end{proposition}
\begin{proof}[Proof of Prop.~\ref{prop:unbounded:parallel:momentum}] This estimate is an outcome of Eq.~\eqref{eq:estimate:Boltzmann:equilibrium}, together with the computation of the momentum by means of Eq.~\eqref{sys:parallel:impilicit:c}.
\end{proof}

\begin{remark}[Collocated versus staggered discretizations]\label{rem:collocated}
  The consistency with the asymptotic equilibria is closely related to the staggered definition of the discrete spatial operators as defined by Eqs.~\eqref{sys:def:discrete:space:opertors}. Using collocated schemes, all the variables are computed on the grid nodes including the momentum and gradient components. This entails the following definitions
  \begin{subequations}
  \begin{gather}
  \begin{multlined}[0.8\textwidth]\label{eq:collocated:momentum}
    \left((n_e u_e)_\parallel\right)^{k+1}_{i,j} - \left((\widetilde{n_e u_e})_\parallel\right)^{k+1}_{i,j} = \\[\jot]
    \qquad - \frac{\Delta t}{\varepsilon_{M,e}} \left( T_e(\widehat{\nabla}^h_\parallel n_e^{k+1})_{i,j} + (n_e)^k_{i,j} (\widehat{\nabla}^h_\parallel \phi^{k+1})_{i,j}  \right)\,,
  \end{multlined}\\
    (\widehat{\nabla}^h_\parallel \phi^h)_{i,j} = \frac{1}{2 \Delta y}\left( \phi_{i,j+1} - \phi_{i,j-1}\right)\,,
  \end{gather}
  yielding 
    \begin{align}
      &T_e(\widehat{\nabla}^h_\parallel n_e^{k+1})_{i,j}  = \frac{T_e}{2}\left(({\nabla}^h_\parallel n_e^{k+1})_{i,j+1/2} + ({\nabla}^h_\parallel n_e^{k+1})_{i,j-1/2}\right) \,,\\
      &\begin{aligned}
    (n_e)^k_{i,j} (\widehat{\nabla}^h_\parallel \phi^{k+1})_{i,j} = &\frac{1}{2} \Big(( \bar{n}_e)^k_{i,j+1/2} ({\nabla}^h_\parallel \phi^{k+1})_{i,j+1/2} \\
      & \qquad + ( \bar{n}_e)^k_{i,j-1/2} ({\nabla}^h_\parallel \phi^{k+1})_{i,j-1/2}\Big) + \mathcal{O} \left( \Delta y^2\right)\,,
      \end{aligned}\\
      &\qquad (\bar{n}_e)^k_{i,j+1/2} = \frac{1}{2}\left((n_e)^k_{i,j+1} +(n_e)^k_{i,j}  \right)\,.
    \end{align}
  \end{subequations}
  The classical three point discretization is conserved for the discretization of second order differential operators within collocated schemes (see for instance \cite{alvarez-laguna_asymptotic_2020}). The estimate stated by Eq.~\eqref{eq:estimate:Bontzmann:balance} is thus preserved but that of the parallel momentum reconstructed by means of Eq.~\eqref{eq:collocated:momentum} deteriorates into
  \begin{equation}\label{eq:collocated:estimate}
    \left\|\frac{1}{\Delta t}\left({(n_e\u_{e,h})^{k+1}_\parallel - (\widetilde{n_e\u_{e,h}})^{k+1}_\parallel}\right)\right\|_\infty \sim \frac{{1}}{\varepsilon_{M,e}} \Big( \left\|R_{n_e}^h \right\|_\infty + \mathcal{O}(\Delta y^2) \Big) \,.
  \end{equation}
The growth rate of the parallel momentum is therefore notably increased. For direct solvers, the norm of the residual relative to that of the right-hand side of the linear system can be expected to approach the precision of computer arithmetic. However, the introduction of the truncation error in Eq.~\eqref{eq:collocated:estimate} significantly de the accuracy of the parallel velocity. The truncation error, $\mathcal{O}(\Delta y^2)$, is amplified by $\eps_{M,e}^{-1}$ when reconstructing the parallel momentum; therefore, the discretization parameter must decrease as $\eps_{M,e} \to 0$ to maintain a finite parallel velocity. Using such three-point discrete second-order operators within a collocated scheme appears incompatible with the asymptotic-preserving property.
\end{remark}


\subsection{An asymptotic preserving scheme in the quasi-neutral and drift limits}
The aim here is to obtain a means of computing the parallel momentum without the amplification of any of the errors. In this aim, an auxiliary variable is introduced, denoted $\kappa^{k+1}$, to rescale the stiff term in the momentum equation and derive an asymptotic preserving scheme. The auxiliary variable satisfies the following balance:

\begin{equation}\label{eq:def:kappa}
  \nabla_\parallel \kappa_e^{k+1} = \frac{1}{\eps_{M,e}} \left(T_e\nabla_\parallel n_e^{k+1} - n_e^k \nabla_\parallel \phi^{k+1} \right)\,.
\end{equation}
Owing to this definition, the following set of discretized equations is derived for the simplified framework consisting of a one dimensional aligned problem.
\begin{definition}[Asymptotic Preserving scheme]\label{def:AP:Scheme}
The Asymptotic Preserving (AP) scheme is defined as follows
  \begin{subequations}\label{sys:def:AP:parallel}
    \begin{align}
      n_{e,h}^{k+1} &= n_{i,h}^{k+1}+ \lambda^2 \Delta^h_\parallel \phi^{k+1}_h\,, \label{sys:def:AP:parallel:a}\\
      -\Delta_\parallel^h \kappa_{e,h}^{k+1} &= (\Delta t)^{-2} \left( n_{e,h}^{k} - n_{e,h}^{k+1}\right) - (\Delta t)^{-1}\nabla_\parallel^h \cdot (\widetilde{n_{e,h}\mathbf{u}_{e,h}})^{k}_\parallel\,,\label{sys:def:AP:parallel:b}\\
      -\nabla_\parallel^h \cdot ( n_{e,h}^k \nabla_\parallel^h \phi^{k+1}_h) &= -\nabla_\parallel^h \cdot ( T_e \nabla_\parallel^h n_{e,h}^{k+1}) + \eps_{M,e} \Delta_\parallel^h\kappa_{e,h}^{k+1} \,,\label{sys:def:AP:parallel:c}
    \end{align}
  with  the parallel momentum being obtained thanks to
    \begin{equation}\label{sys:def:AP:parallel:d}
      (n_{e,h}\mathbf{u}_{e,h})^{k+1}_\parallel= (\widetilde{n_{e,h}\mathbf{u}_{e,h}})^{k}_\parallel - \Delta t \nabla_\parallel^h \kappa_{e,h}^{k+1} \,.
    \end{equation}
  \end{subequations}
\end{definition}
\begin{proposition}(Well posed-ness of the AP-scheme in the asymptotic regimes)\label{Prop:Wellposedness:AP}
  Under the assumptions of Hyp.~\ref{hyp:ellipticity:parallel}, the AP scheme defined by Eqs.~\eqref{sys:def:AP:parallel} is well posed in the limit $\lambda\to 0$, $\varepsilon_{M,e}\to 0$ as well as in the combined limit $(\lambda,\eps_{M,e}) \to 0$.
\end{proposition}

\begin{proof}[Proof of Prop.~\ref{Prop:Wellposedness:AP}]
  The matrix of the linear system \eqref{eq:AP:linear:system:parallel} is triangular for $\lambda=0$, owing to the ellipticity assumptions (Hyp.~\ref{hyp:ellipticity:parallel}), the system is invertible.

  In the combined quasi-neutral and drift limits $(\lambda,\eps_{M,e})\to 0$, the electric potential is obtained thanks to Eq.~\eqref{sys:def:AP:parallel:c}. The density is deduced from Eq.~\eqref{sys:def:AP:parallel:a} entailing the matching of the electron and ion densities. Finally, the auxiliary variable is carried out thanks to Eq.~\eqref{sys:def:AP:parallel:b}.
\end{proof}

The first three equations of the AP-scheme are coupled and yield the resolution of a linear system that may be recast into the following block linear system
\begin{equation}\label{eq:AP:linear:system:parallel}
  \begin{gathered}
  \begin{pmatrix}
     \Id & 0 & -\lambda^2 \Delta_\parallel^h  \\
    (\Delta t)^{-2} \Id &-\Delta_\parallel^h& 0 \\
     \Delta_{T_e,\parallel}^h & -\eps_{M,e} \Delta_\parallel^h & -\Delta_{n_e,\parallel}^h 
  \end{pmatrix}\begin{pmatrix}
    n_{e,h}^{k+1} \\ \kappa^{k+1}_{e,h} \\ \phi^{k+1}_h 
  \end{pmatrix} 
  = \begin{pmatrix}
    n_i^{k+1} \\ (\Delta t)^{-2}  \widetilde{n}_{e,h}^{k}  \\ 0  
  \end{pmatrix} + \begin{pmatrix}
    R^h_{\phi} \\ R^h_{\kappa_e}    \\  R^h_{n_e}
  \end{pmatrix}\,.\\
\end{gathered}
\end{equation}
In Eq.~\eqref{eq:AP:linear:system:parallel}, the vector $R^h_{\phi}$, $R^h_{n_e}$ and $R^h_{\kappa_e}$ are the residual issued from the numerical resolution of the linear system.

The AP-scheme enjoys the same property as the implicit scheme regarding the quasi-neutral limit.
\begin{proposition}[Consistency of the AP scheme with the quasineutrality constraint]\label{prop:AP:consistency:QN}
  In the quasi-neutral limit $\lambda = 0$ the charge density, carried out by the AP-scheme defined Eq.~\eqref{eq:AP:linear:system:parallel}, is controlled by the linear system residual $R^h_\phi$
  \begin{equation}\label{eq:estimate:charge:density:AP}
    \left\| n_{e,h}^{k+1} - n_{i,h}^{k+1} \right\|_\infty= \left\| R_\phi^h \right\|_\infty.
  \end{equation}  
\end{proposition}
The proof of Prop.~\ref{prop:AP:consistency:QN} is similar to that of Prop.~\ref{prop:quasineutral:limit}, the first line of both schemes being the same.
\begin{proposition}[Discrete consistency of the AP-scheme with the Boltzmann equilibrium]\label{prop:}
  Provided that the elliptic operators are issued from the discrete divergence and gradient operators as defined by Eqs.~\eqref{sys:def:discrete:space:opertors} as well as a zero flux condition, for both the density and the electric potential, at one end of the magnetic field lines, the Boltzmann equilibrium issued from Eq.~\eqref{eq:AP:linear:system:parallel} in the limit $\eps_{M,e} \to 0$ satisfies 
  \begin{equation}
    \left\| n_{e,h}^k \nabla_\parallel \phi^{k+1}_h  - T_e \nabla_\parallel^h n_{e,h}^{k+1}\right\|_\infty  \sim  \left\| R^h_{n_e}\right\|_\infty \,.
  \end{equation}
\end{proposition}

\begin{proof} The estimate is obtained by setting $\varepsilon_{M,e}=0$ into the third block line of the system matrix and following the arguments of Prop.~\ref{prop:Bolzmann:equilibrium} proof.
\end{proof}
The computation of the parallel momentum genuinely provides a means of computing the parallel momentum for any value of the parameter $\eps_{M,e}\geq 0$.
\begin{proposition}\label{prop:bounded:parallel:momentum}
  The parallel momentum issued from the AP scheme defined by Eqs.~\eqref{sys:def:AP:parallel} remains finite in the drift limit.
\end{proposition}
\begin{remark}
  Contrary to Eq.~\eqref{sys:parallel:impilicit:c} used in the implicit scheme, the property of the parallel velocity to remain bounded in the drift limit $\varepsilon_{M,e}\to 0$ is preserved by Eqs.~\eqref{sys:def:AP:parallel} for discrete second order differential operators that are not issued from staggered discretizations as defined by Eqs.~\eqref{sys:def:discrete:space:opertors}. This allows the use of collocated schemes contrary to formulation \eqref{sys:parallel:impilicit:c} embedded in the implicit scheme (See Rem.~\ref{rem:collocated}).
\end{remark}

It is worth mentioning that the size of the system issued from the AP-scheme is significantly larger than that of the implicit scheme one. Considering the only evolution of the electrons, the introduction of the auxiliary variable increases the number of unknowns from 2 to 3. Restoring the perpendicular dynamics as well as the evolution of the ions, which is chosen implicit, the system writes
\begin{subequations}\label{sys:AP:total}
  \begin{align}
  &-\lambda^2\Delta^h\phi_h^{k+1} =n_{i,h}^{k+1} - n_{e,h}^{k+1}\,,\label{sys:AP:total:a}\\
  &-\varepsilon_{M,e} \Delta_\parallel^h \kappa_{e,h}^{k+1} = \nabla_\parallel^h\cdot \left(n_{e,h}^h \nabla_\parallel^h \phi_h^{k+1} \right) - T_e \Delta_\parallel^h n_{e,h}^{k+1}\,,\label{sys:AP:total:b}\\
  &\begin{multlined}[0.85\textwidth]
   \quad n_{e,h}^{k+1} -  \frac{T_e \Delta t^2}{\varepsilon_{M,e}} \nabla_\perp^h \cdot\left(\mathbb{A}_{e}^\perp\nabla_\perp^h n_e^{k+1}\right) = \Delta t^2 \Delta^h_\parallel \kappa_{e,h}^{k+1} + \widetilde{n}_{e,h}^k \\
     +\frac{\Delta t^2}{\varepsilon_{M,e}} \nabla_\perp^h\cdot \left(n_{e,h}^k  \mathbb{A}_{e}^{\perp}  \nabla^h_\perp\phi_h^{k+1}\right)\,,\label{sys:AP:total:c}
    \end{multlined}\\
    &-\varepsilon_{M,i} \Delta_\parallel^h \kappa_{i,h}^{k+1} = \nabla_\parallel^h\cdot \left(n_{i,h}^h \nabla_\parallel^h \phi_h^{k+1} \right) - T_i \Delta_\parallel^h n_{i,h}^{k+1}\,,\label{sys:AP:total:d}\\
    &\begin{multlined}[0.85\textwidth]
     \quad n_{i,h}^{k+1} -  \frac{T_i \Delta t^2}{\varepsilon_{M,i}} \nabla_\perp^h \cdot\left(\mathbb{A}_{i}^\perp\nabla_\perp^h n_i^{k+1}\right) = \Delta t^2 \Delta^h_\parallel \kappa_{i,h}^{k+1} + \widetilde{n}_{i,h}^k \\
       +\frac{\Delta t^2}{\varepsilon_{M,i}} \nabla_\perp^h\cdot \left(n_{i,h}^k  \mathbb{A}_{i}^{\perp}  \nabla^h_\perp\phi_h^{k+1}\right)\,.\label{sys:AP:total:e}
      \end{multlined}
\end{align}
These five coupled equations provide  the five unknowns $n_{e,h}^{k+1}$, $n_{i,h}^{k+1}$, $\kappa_{e,h}^{k+1}$, $\kappa_{i,h}^{k+1}$ and $\phi^{k+1}_h$, the (electron) momentum is reconstructed thanks to
\begin{equation}
  \begin{aligned}
    (n_e\mathbf{u}_{e,h})^{k+1}_\parallel &=  (\widetilde{n_{e,h}\mathbf{u}_{e,h}})^{k}_\parallel -\Delta t \nabla_\parallel^h \kappa_{e,h}^{k+1}\,,\\
    (n_e\mathbf{u}_{e,h})^{k+1}_\perp &= \mathbb{A}_e^\perp \left[  (\widetilde{n_{e,h}\mathbf{u}_{e,h}})^{k}_\perp - \frac{\Delta t}{\varepsilon_{M,e}} \left(T_e \nabla_\perp^h n_{e,h}^{k+1} +  n_{e,h}^k \nabla_\perp^h \phi_h^{k+1}\right) \right]\,,\\
    (\widetilde{n_{e,h}\mathbf{u}_{e,h}})^{k}_\perp&={(n_e\mathbf{u}_{e,h})^{k}_\perp} - \Delta t\nabla_\perp\cdot(n_{e,h} \mathbf{u}_{e,h}   \otimes\mathbf{u}_{e,h})^k_\perp\,.
    \end{aligned}
\end{equation}
\end{subequations}
A similar set of equations is used for the ions. The coupled system may be recast under the following block matrix form:
\begin{equation*}
  \begin{pmatrix}
  M_{\phi,\phi} &    M_{\phi,n_e}  & 0    & M_{\phi,n_i}   &0\\
       M_{n_e,\phi} &     M_{n_e,n_e}  &   M_{n_e,\kappa_e} &    0 & 0 \\
       M_{\kappa_e,\phi} &     M_{\kappa_e,n_e} &     M_{\kappa_e,\kappa_e} &     0 & 0 \\
       M_{n_i,\phi} &    0 & 0 &    M_{n_i,n_i}  &   M_{n_i,\kappa_i} \\
       M_{\kappa_i,\phi}   & 0 & 0 &    M_{\kappa_i,n_i}  &   M_{\kappa_i,\kappa_i}
  \end{pmatrix}
  \begin{pmatrix}
  \phi_h^{k+1} \\ n_{e,h}^{k+1} \\ \kappa_{e,h}^{k+1} \\ n_{i,h}^{k+1} \\ \kappa_{i,h}^{k+1}
  \end{pmatrix}
  =
  \begin{pmatrix}
  F_\phi \\ F_{n_e} \\ F_{\kappa_e} \\ F_{n_i} \\ F_{\kappa_i}
  \end{pmatrix}.
  \end{equation*}
 \subsection{Enhancements of the AP scheme}
Several comments can be stated at this point. 

First, the implicit coupling between Poisson's equation and the ion system implemented in Eqs.~\eqref{sys:AP:total} is not mandatory. An implicit coupling with the electron is sufficient to cope with the degeneracy of Poisson's equation in the quasi-neutral limit. With an explicit discretization of the electric force within the ion system, the computation of the ion density Eqs.~(\ref{sys:AP:total:d}--\ref{sys:AP:total:e}) may be decoupled from that of the electron and the electrostatic potential. 

Second, though the electrons are in the drift approximation, the ion to electron mass ratio is large enough for the ion Mach number to be close to one. The variable $\kappa_i$ may thus be discarded in the formulation hence reducing the size of the system attached to the ions. 

Third, block preconditioners designed for similar problems, as recently introduced in \cite{li_block_2024}, enable a significant improvement in the efficiency of linear system solvers for such block matrices compared to sparse direct solvers. This block preconditionners are used together iterative solvers and particularly efficient for three-dimensional applications. This emphasizes the advantage of a numerical method free from the residual amplification as demonstrated by the herein introduced AP-scheme.

Finally, it is important to note that the auxiliary variable can be introduced as a post-processing step following the implicit scheme, allowing for an approximation of the parallel momentum without amplifying the residual of the linear system. For simplicity, this alternative scheme is presented within the context of Def.~\ref{def:AP:Scheme}. As in the implicit scheme, the electron density and electrostatic potential are computed by solving the coupled Poisson and reformulated continuity equations:
\begin{subequations}\label{sys:def:AP:alternative:parallel}
  \begin{align}
    &n_{e,h}^{k+1} = n_{i,h}^{k+1}+ \lambda^2 \Delta^h_\parallel \phi^{k+1}_h\,, \label{sys:def:AP:alternative:parallel:a}\\
    &
    -\nabla_\parallel^h \cdot ( n_{e,h}^k \nabla_\parallel^h \phi^{k+1}_h) = {-}\nabla_\parallel^h \cdot ( T_e \nabla_\parallel^h n_{e,h}^{k+1}) + \frac{\eps_{M,e}}{\Delta t^2} \left(\widetilde{n}_{e,h}^{k} - n_{e,h}^{k+1}\right) \,,\label{sys:def:AP:alternative:parallel:c}
  \end{align}
  where
  \begin{equation}\label{sys:def:AP:alternative:parallel:bis}
     \widetilde{n}_{e,h}^{k} = n_{e,h}^{k} - {(\Delta t)^{-1}}\nabla_\parallel^h \cdot (\widetilde{n_{e,h}\mathbf{u}_{e,h}})^{k}_\parallel\,.
  \end{equation}
  The parallel momentum computation is then decomposed into two steps. First the of the auxiliary variable is carried out thanks to the electron density and electric potential issued from Eqs.~(\ref{sys:def:AP:alternative:parallel:a}-\ref{sys:def:AP:alternative:parallel:bis}), then it is used to define an approximation free from any amplification of the residual:
  \begin{align}\label{sys:def:AP:alternative:parallel:d}
    {-}\Delta_\parallel^h \kappa_{e,h}^{k+1} &= {(\Delta t)^{-2}} \left( \tilde{n}_{e,h}^{k} - n_{e,h}^{k+1}\right)\,,\\
    (n_{e,h}\mathbf{u}_{e,h})^{k+1}_\parallel&= (\widetilde{n_{e,h}\mathbf{u}_{e,h}})^{k}_\parallel - \Delta t \nabla_\parallel^h \kappa_{e,h}^{k+1} \,.
  \end{align}
\end{subequations}
This two-step formulation of the AP-scheme improves computational efficiency by reducing the size of the linear system. Instead of solving a single system with $3(N_x\times N_y)$ rows, the scheme first solves a system with $2 (N_x\times N_y)$ rows for the coupled computation of the electric potential and density, followed by a system with $(N_x\times N_y)$ rows for the auxiliary variable. The overhead, compared to the implicit is therefore limited to the resolution of an additional elliptic problem.

These alternative formulations of the proposed scheme as well as the development of efficient linear system solvers  shall be investigated in future works.

\section{Numerical investigations}
\label{sec:results}
\subsection{Introduction}
In this section, the abilities of the numerical scheme proposed in the paper are numerically assessed. The first test is devoted to investigate the capabilities if the scheme to reproduce the confinement of the plasma by a strong magnetic field. This test case highlights the difficulty of the implicit scheme to reproduce the plasma confinement due to the amplification of errors in the low Mach regime. This weakness is corrected thanks to the AP scheme. Second, the influence of neutral particles on the plasma dynamics is investigated in order to decrease the total amount of energy received by the target. Third, a classical sheath problem is considered to numerically experience the ability of the scheme to deal with local quasi-neutrality breakdown.

\subsection{Confinement with a strong magnetic field}\label{sec:num_low_Mach}
\begin{figure}[htbp]
\begin{center}
\subfigure[1D initial densities\label{fig:densityinitial:a}]{
\includegraphics[width= 0.17\textwidth]{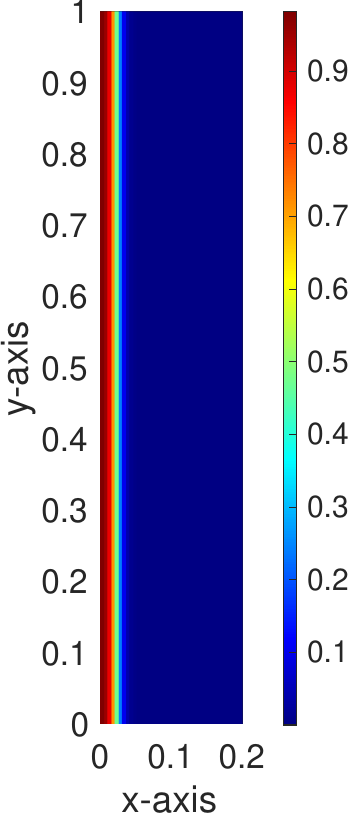}}\hspace*{0.1\textwidth}
\subfigure[Perturbation of the 1D initial densities\label{fig:densityinitial:b}]
  { \includegraphics[width= 0.185\textwidth]{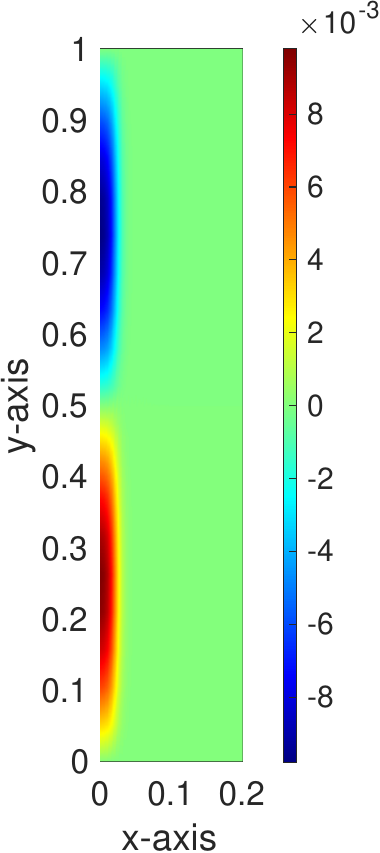}}
\end{center}
\caption{Plasma confinement: Initial electron and ion densities for the 1D steady state data and the perturbation of the steady state.\label{fig:densityinitial}}
\end{figure}

This test case is two-dimensional and related to the operating of the HIT-PSI device \cite{wang2023}. The aim here, is to investigate the property of the schemes to preserve stationary solutions and assess numerically the stability to perturbations of these solutions. The characteristic scales attached to these computations are gathered in Tab.~\ref{tab:physics} (See Appendix~\ref{sec:RQ}), the plasma is evolving under in $\Omega=[0,0.2]\times[0,1]$. The electron temperature is usually higher than the ion one, yielding the following setting $T_i=0.05$ and $T_e=2$.
The initial density is defined by
\begin{equation*}
n_\alpha = \frac{1}{2} \Big( 1+ f_n(x) +  10^{-4}\big(1 - f_n(x)\big) \Big)\,, \quad f_n(x)=\tanh\big(10^2(0.02-x)\big),\quad \alpha\in\{i,e\},
\end{equation*}
which defines a one-dimensional initial data illustrated in Figure~\ref{fig:densityinitial}(a). 
The ions and electrons have the same initial velocity $\mathbf{u}_\alpha = (0,1,0)^T$, $\alpha\in\{i,e\}$. For all scalar variables, axisymmetric boundary conditions are used at $x=0$, Neumann boundary conditions are considered at $x=0.2$, and Neumann boundary conditions are used at $y=0$ and $y=1$, except Dirichlet boundary condition is imposed at $y = 1$ for $\phi$ and $\kappa_\alpha$. The mesh size is $N_x\times N_y = 40\times20$, while the time step is $\Delta t = 10^{-3}$. For these computations, the Debye length is not resolved by the mesh step in regions close to the axis $x=0$ where the plasma density is the largest ($\lambda \sim 10^{-6}$). Conversely, near the boundary $x=0.02$ the plasma density drop is significant, and the Debye length well resolved by the mesh step.

\begin{figure}[htbp]
\begin{center}
\subfigure[Implicit Scheme]{\includegraphics[width=  0.4\textwidth]{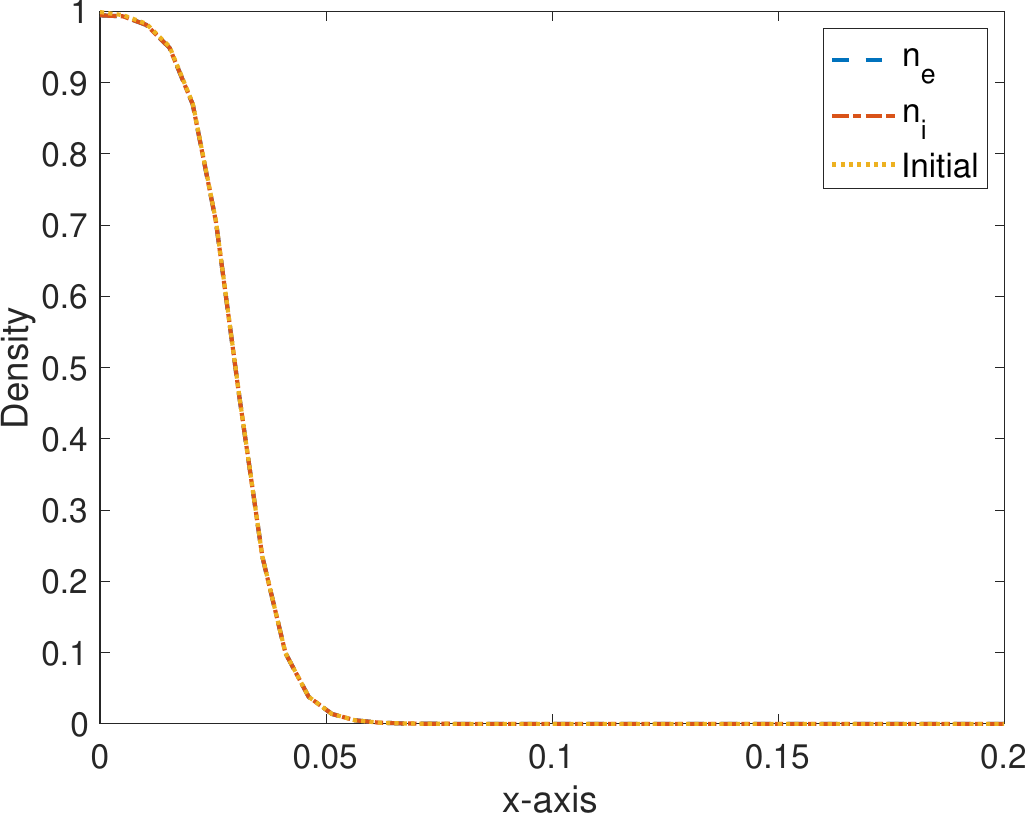}}\hspace*{0.05\textwidth}
\subfigure[AP Scheme]{\includegraphics[width=  0.4\textwidth]{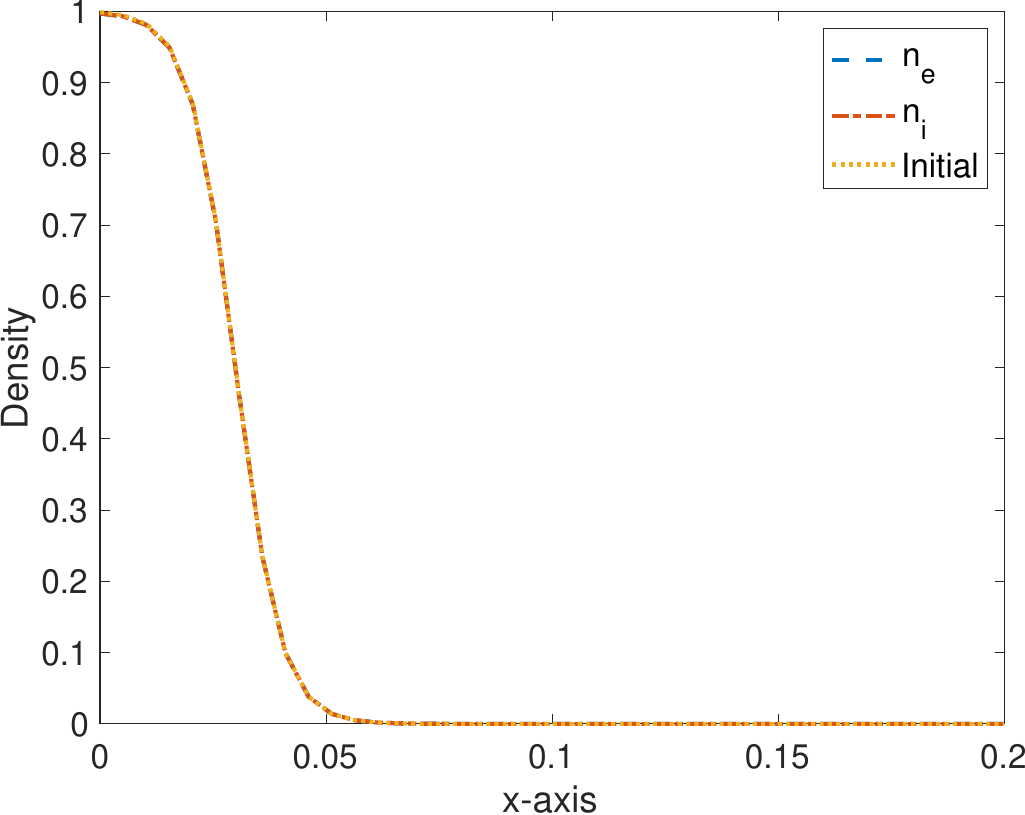}}
\end{center}
\caption{Electron and ion densities (slices at $y=0.5$) as functions of $x$ at initial time and at time t=10.\label{fig:test_confinement_density}}
\end{figure}

\begin{figure}[htbp]
    \begin{center}
    \subfigure[IS: $(n_\alpha u_\alpha)_x$]{\includegraphics[width= 0.3\textwidth]{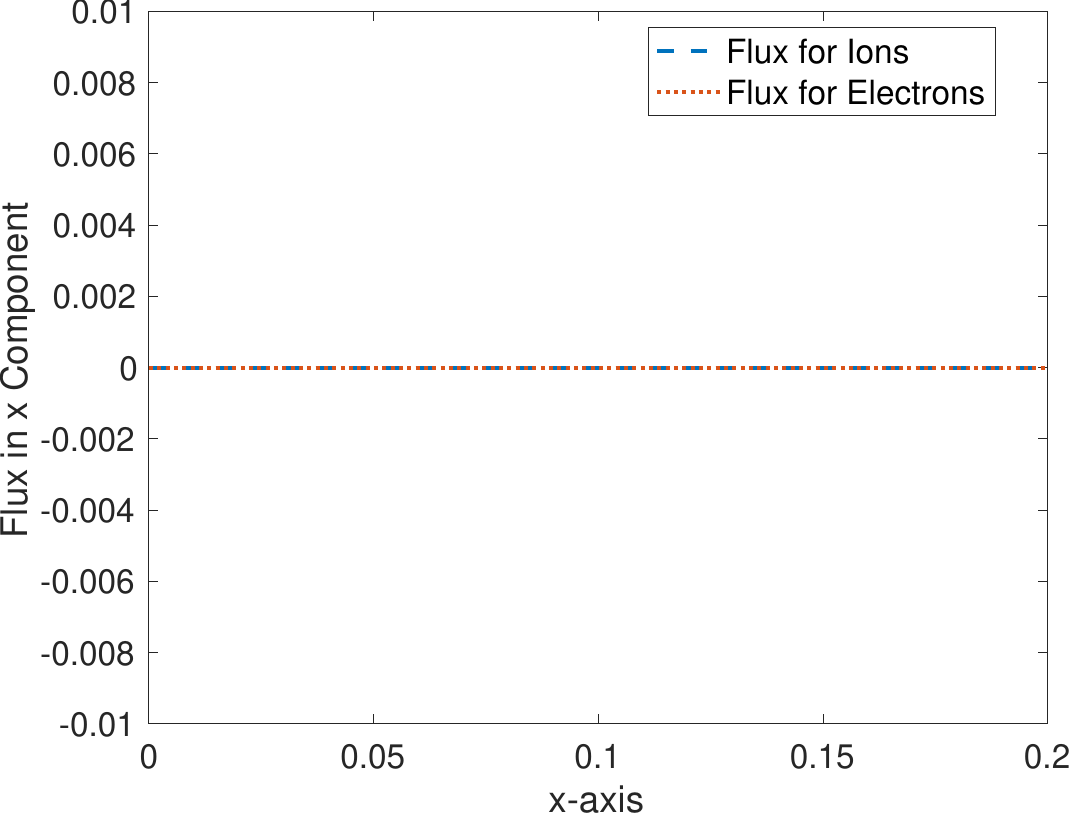}}%
    \hspace*{0.04\textwidth}%
    \subfigure[IS: $(n_\alpha u_\alpha)_y$]{\includegraphics[width= 0.3\textwidth]{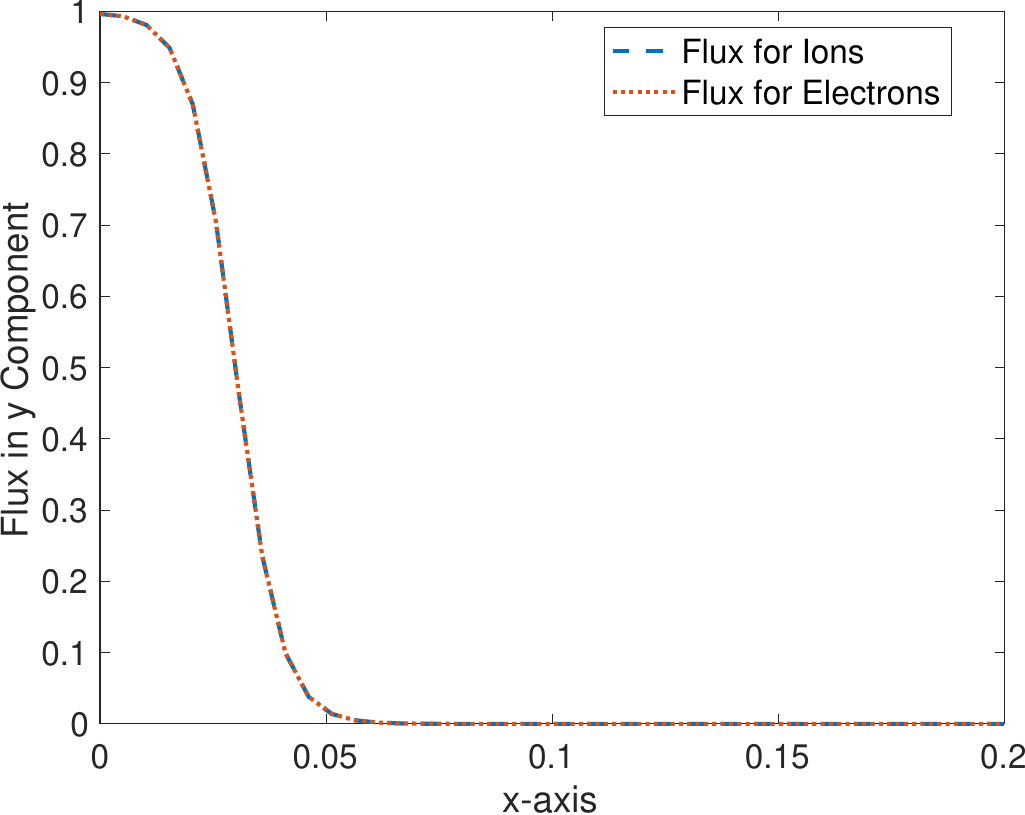}}%
    \hspace*{0.04\textwidth}%
    \subfigure[IS: $(n_\alpha u_\alpha)_z$]{\includegraphics[width= 0.3\textwidth]{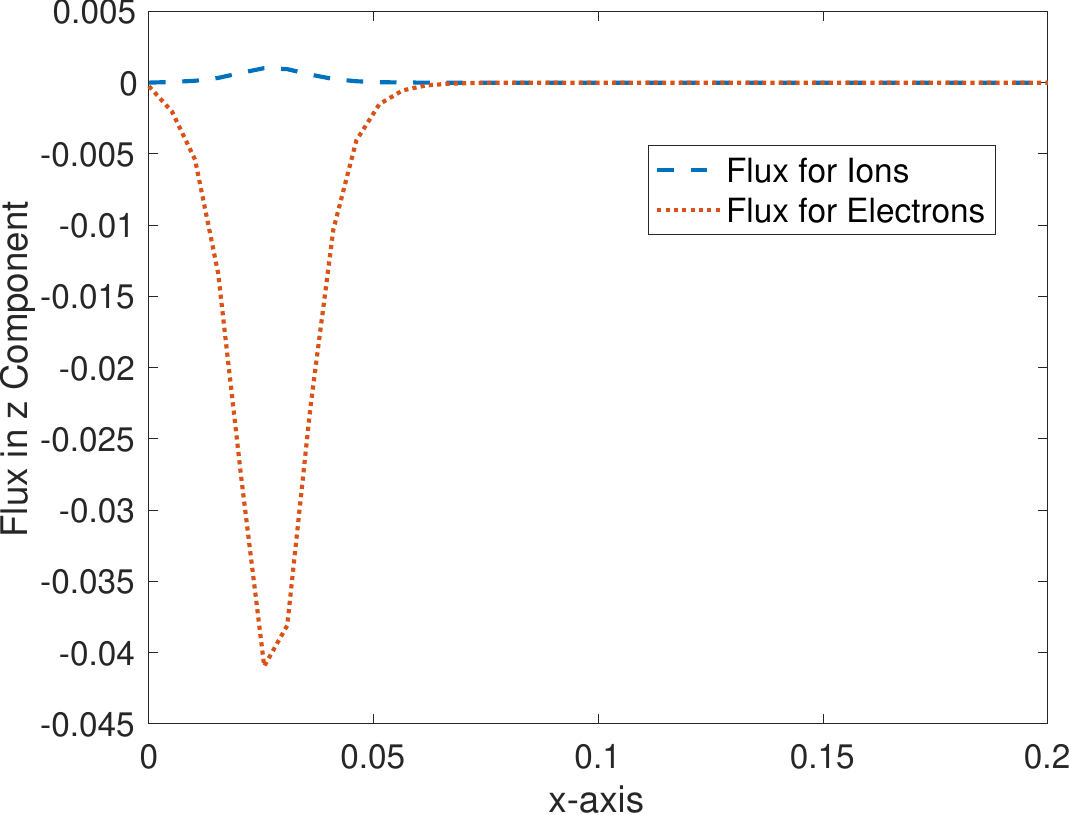}}

    \subfigure[AP: $(n_\alpha u_\alpha)_x$]{\includegraphics[width= 0.3\textwidth]{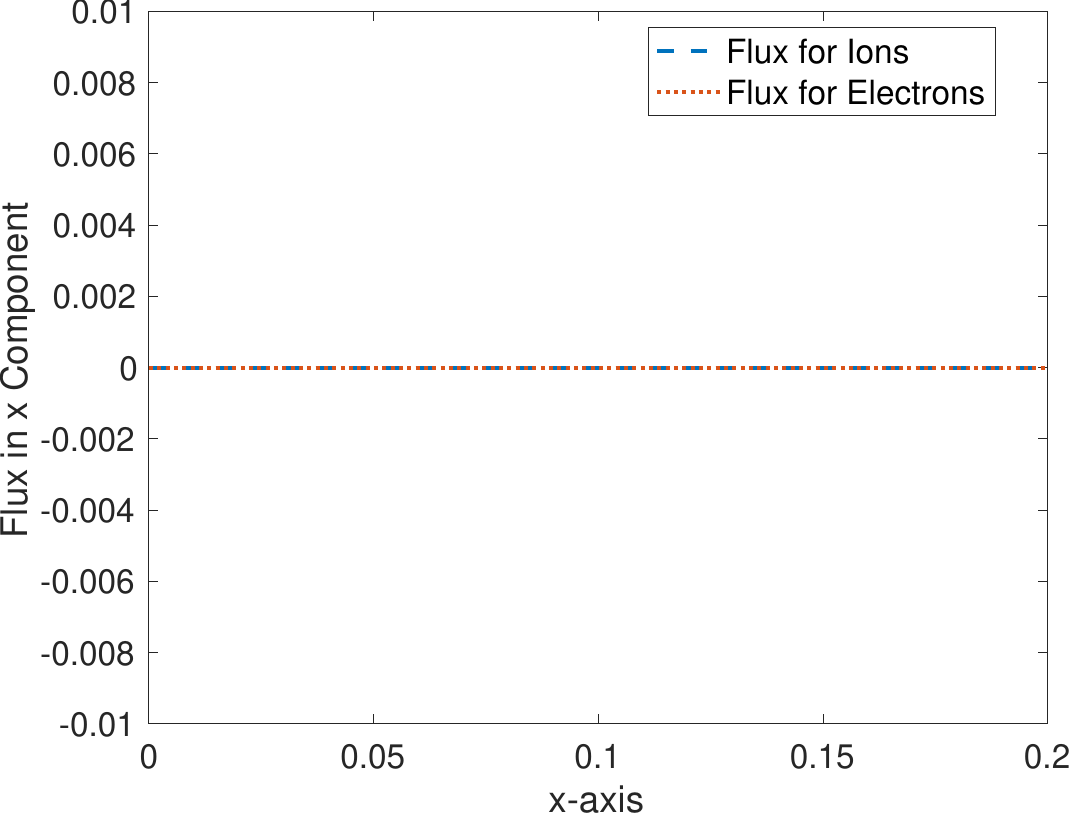}}%
    \hspace*{0.04\textwidth}%
    \subfigure[AP: $(n_\alpha u_\alpha)_y$]{\includegraphics[width= 0.3\textwidth]{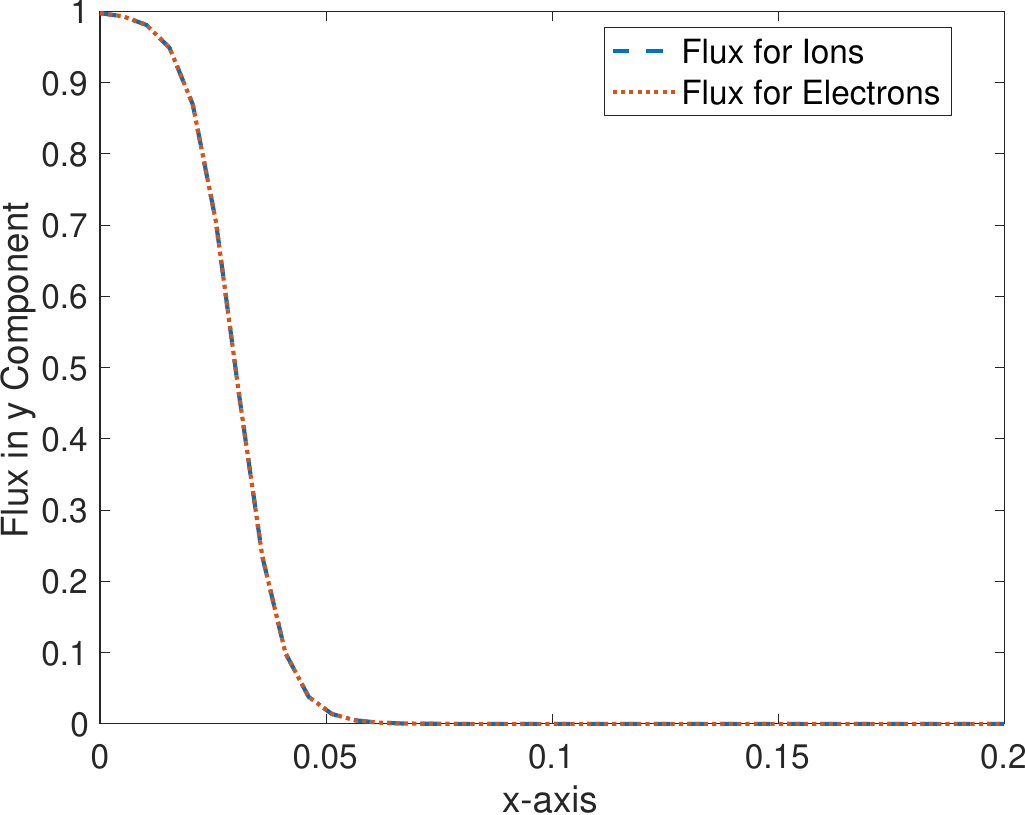}}%
    \hspace*{0.04\textwidth}%
    \subfigure[AP: $(n_\alpha u_\alpha)_z$]{\includegraphics[width= 0.3\textwidth]{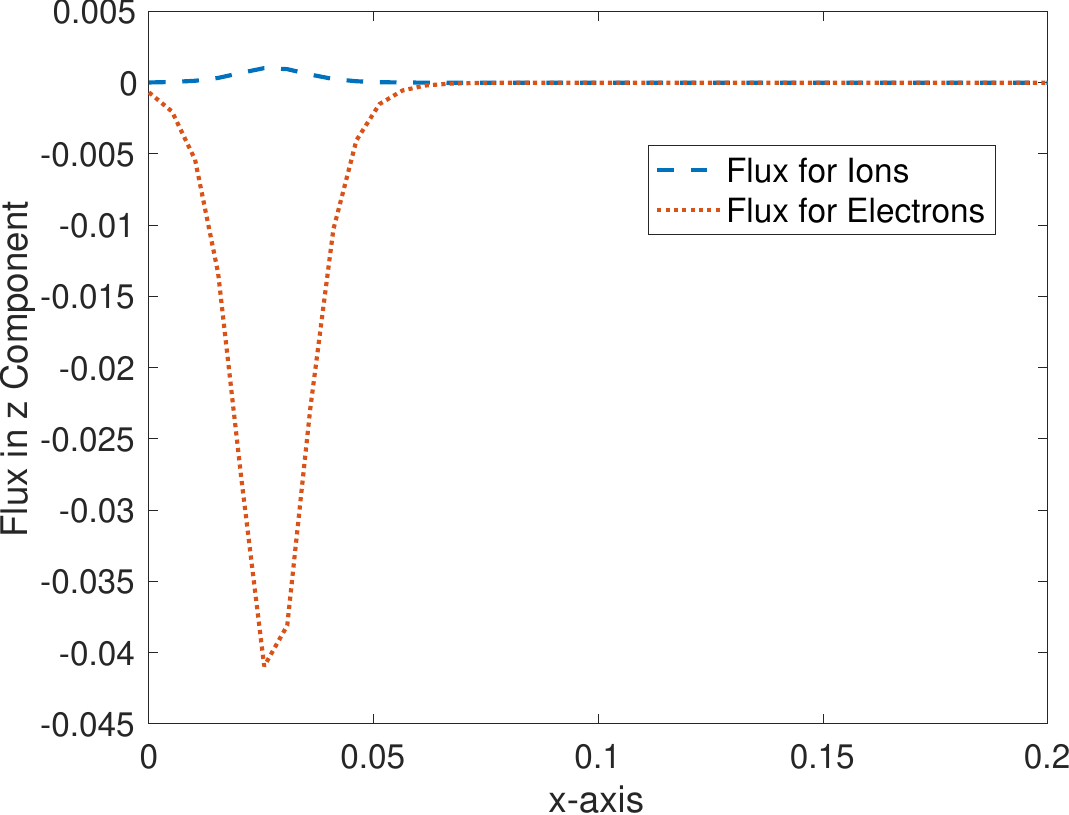}}
\end{center}
\caption{Unperturbed 1D initial data: Ion and electron momentum components (slices at $y=0.5$) computed by the implicit (IS) and AP schemes (AP), at (dimensionless) time t=10 as functions of $x$.\label{fig:test_confinement_flux}}
\end{figure}

The plasma evolution as computed by either the implicit or the asymptotic-preserving schemes is represented on Figs.~\ref{fig:test_confinement_density}-\ref{fig:test_confinement_flux}. 
From Fig.~\ref{fig:test_confinement_density}, the plasma confinement is observed to be well reproduced by both schemes. Indeed, despite the presence of a pressure gradient along the $x$-direction, the plasma density profile remains unchanged as time evolves.
The electron and ion fluxes are plotted on Fig.~\ref{fig:test_confinement_flux}. 
The $y$-component (aligned direction) of the flux is very similar to the profile of the density, which means the velocity in the aligned direction remains constant and equal to the initial one.
Under this condition of intense magnetization of the plasma the drift approximation of the perpendicular velocities prevails as stated by Eq.~\eqref{eq:perpendicular:drifts}. The pressure gradient as well as the electric field are aligned with the $x$-axis defining a zero $x$-component for the velocity but a non-zero component in the $z$-direction.
These observations are recovered from the simulations carried out by either the implicit or the AP scheme, the radial component ($x$-direction) of the flux being totally negligible (see Fig.~\ref{fig:test_confinement_flux}). The poloidal component ($z$-direction) of the momentum is concentrated in regions of large pressure gradients. 

 This is totally in line with the scaling issued from the momentum equation. Both perpendicular components are created in response to the pressure gradient oriented along the $x$-axis with a drift velocity magnitude controlled by the coefficients of the mobility matrix related to the perpendicular direction (see Eq.~\eqref{eq:perpendicular:mobility:matrix}). This yields a $z$-component scaling as $\Omega_\alpha^{-1} M_\alpha^2\sim 10^{-2}$.
 Globally, the dynamics along the perpendicular directions (the radial and the poloidal directions) is expected to be very weak compared to the one in the aligned direction, a feature well reproduced by both schemes.

To investigate further the stability properties of the schemes, a perturbation of the density is added to the initial  condition. This perturbation, illustrated on Fig.~\ref{fig:densityinitial}(b), is defined, for any of the particle density $n_\alpha$, as
\begin{equation*}
\delta n_\alpha = n_\alpha \times (10^{-2}\sin(2\pi y)),\qquad \alpha\in\{i,e\}\,.
\end{equation*}

\begin{figure}[htbp]
  \begin{center}
  \subfigure[AP: $n_\alpha$, $t=0.0$]{\includegraphics[width= 0.3\textwidth]{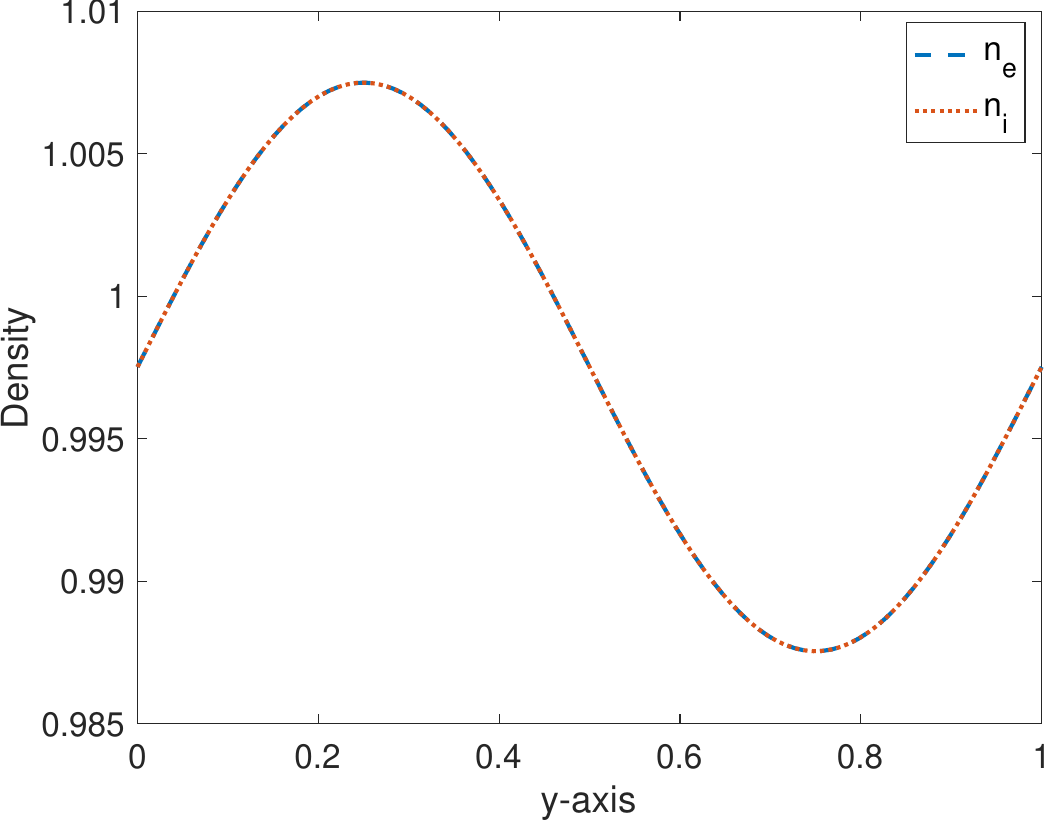}}%
  \hspace*{0.04\textwidth}%
  \subfigure[AP: $n_\alpha$, $t=1$]{\includegraphics[width= 0.3\textwidth]{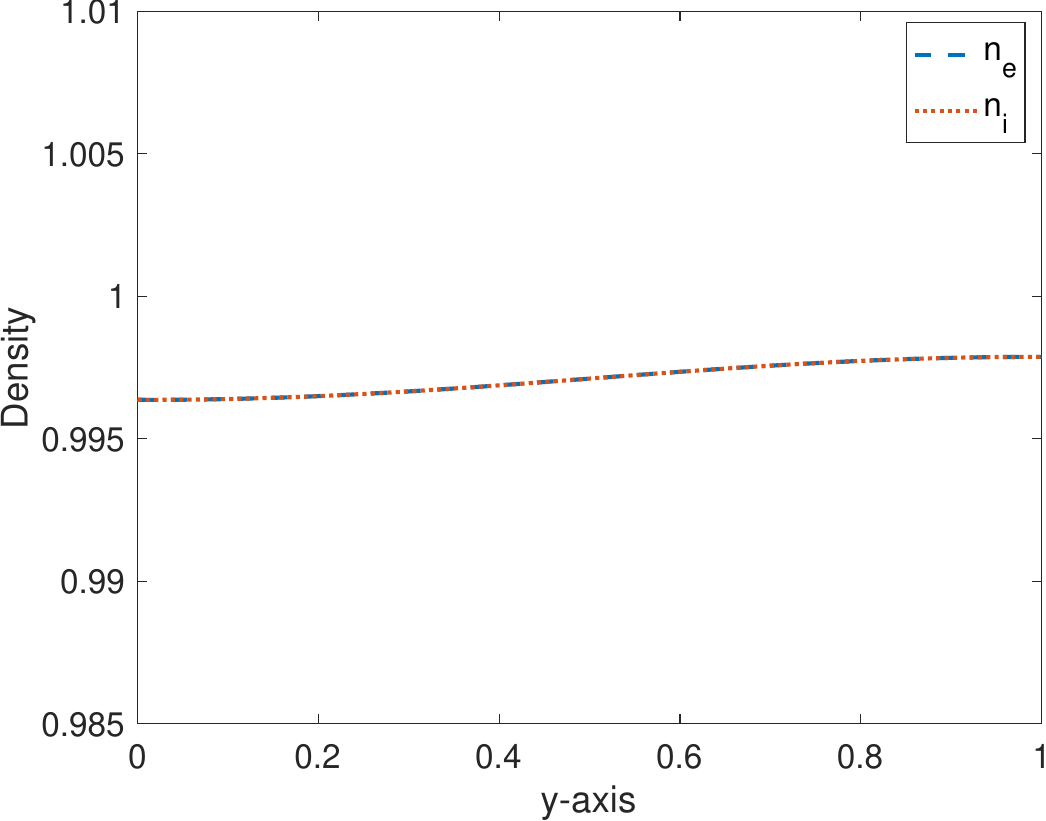}}%
  \hspace*{0.04\textwidth}
  \subfigure[AP: $n_\alpha$, $t=10$]{\includegraphics[width= 0.3\textwidth]{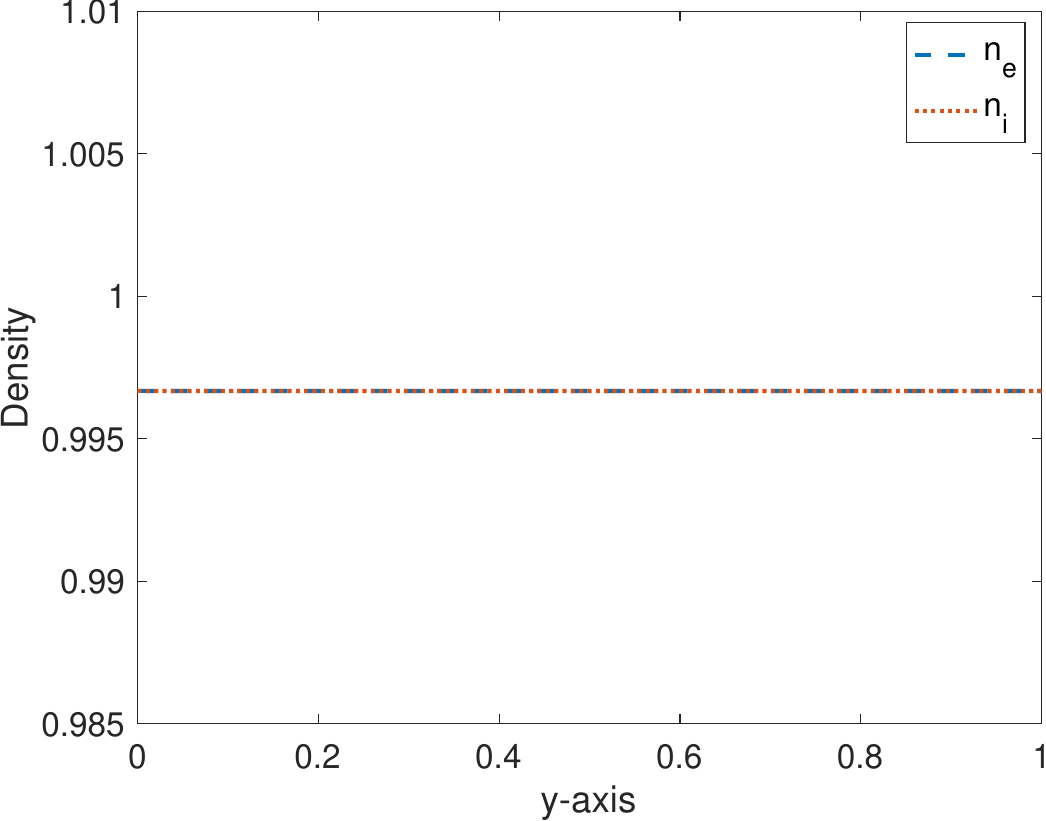}}
  \subfigure[AP: $\phi$, $t=0.0$]{\includegraphics[width= 0.3\textwidth]{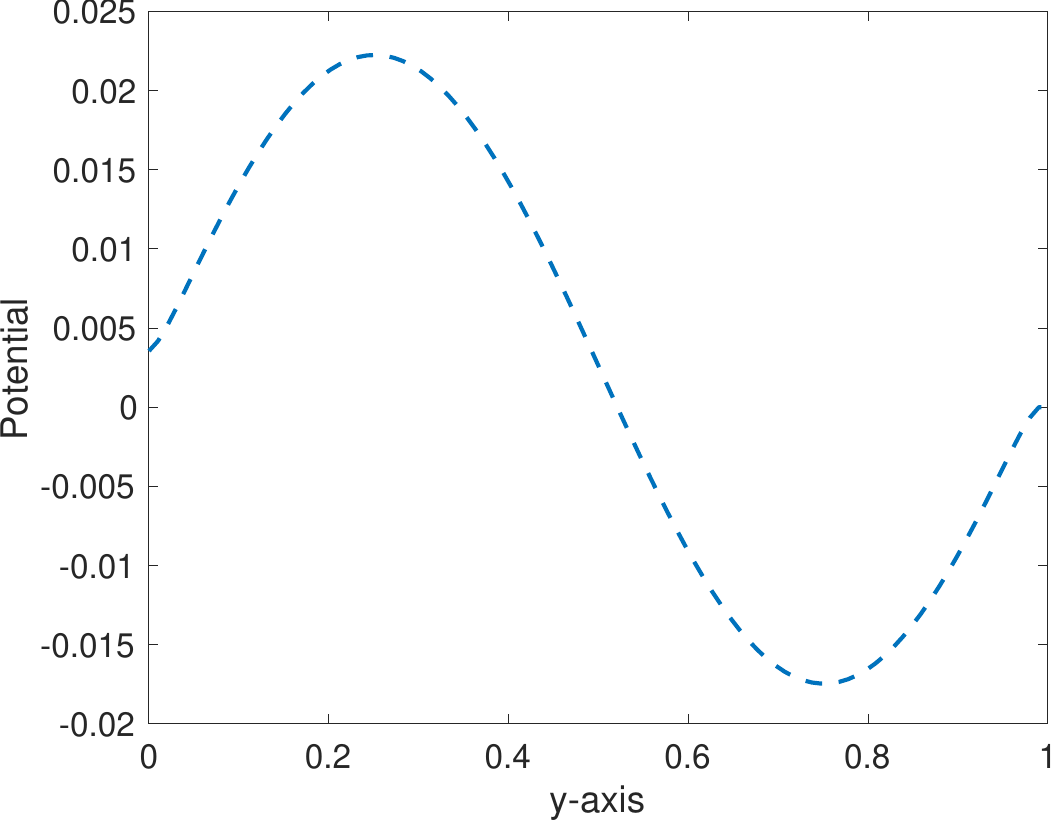}}%
  \hspace*{0.04\textwidth}%
  \subfigure[AP: $\phi$, $t=1$]{\includegraphics[width= 0.3\textwidth]{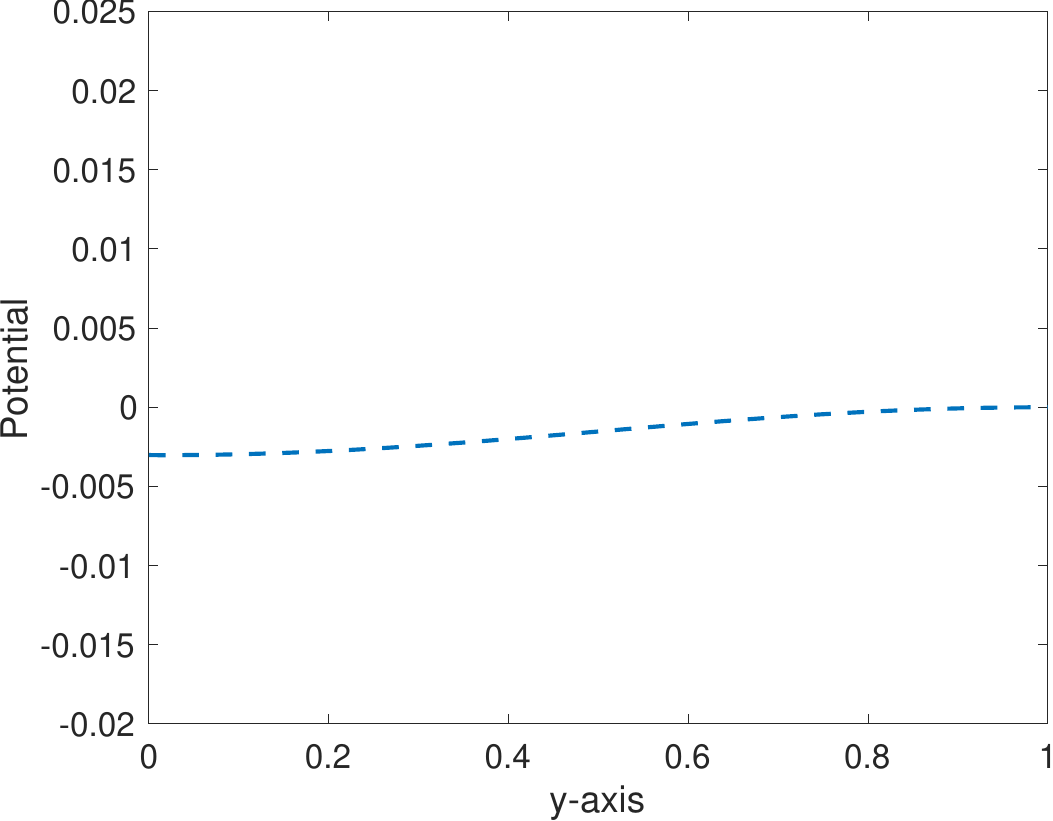}}%
  \hspace*{0.04\textwidth}
  \subfigure[AP: $\phi$, $t=10$]{\includegraphics[width= 0.3\textwidth]{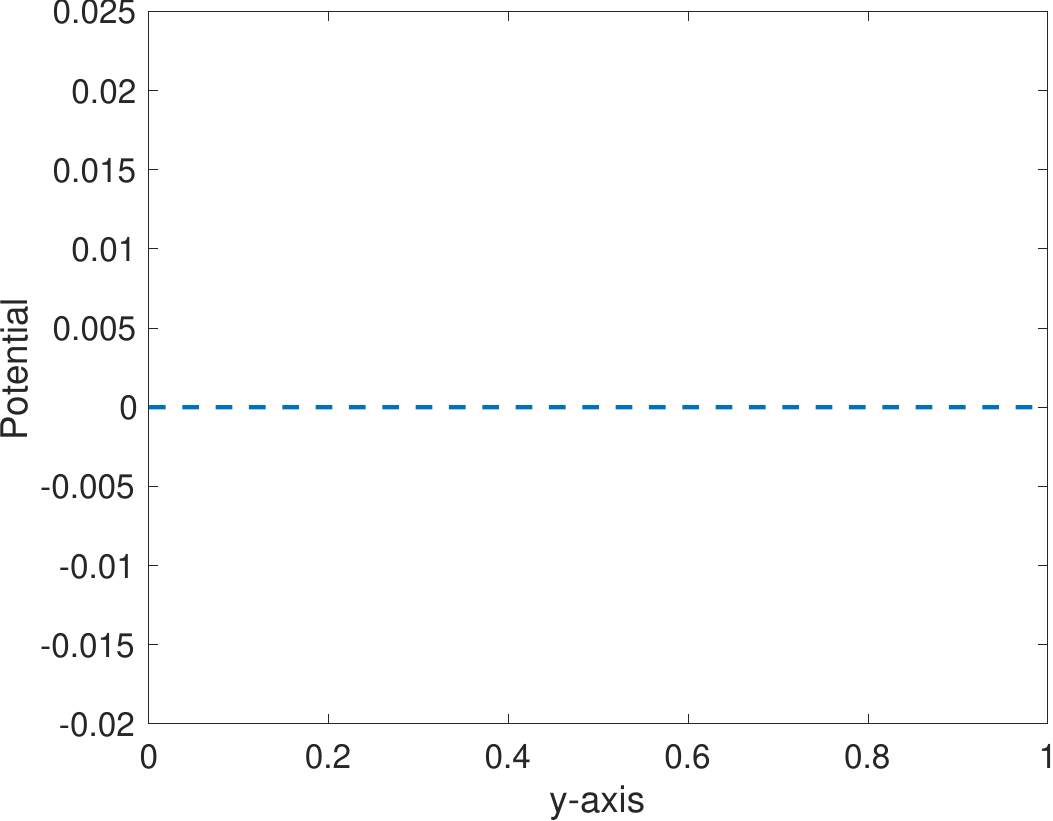}}
  \subfigure[AP: $(n_\alpha \mathbf{u}_\alpha)_{y}$, $t=0$]{\includegraphics[width= 0.3\textwidth]{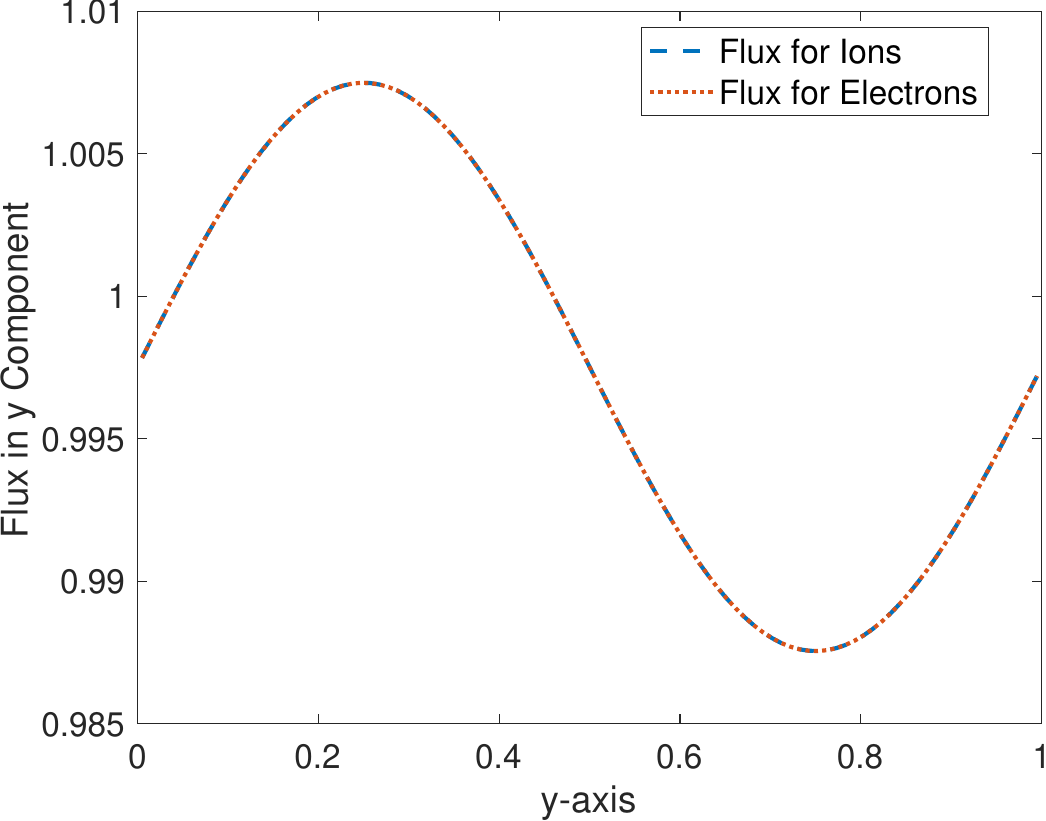}}%
  \hspace*{0.04\textwidth}%
  \subfigure[AP: $(n_\alpha \mathbf{u}_\alpha)_{y}$, $t=1$]{\includegraphics[width= 0.3\textwidth]{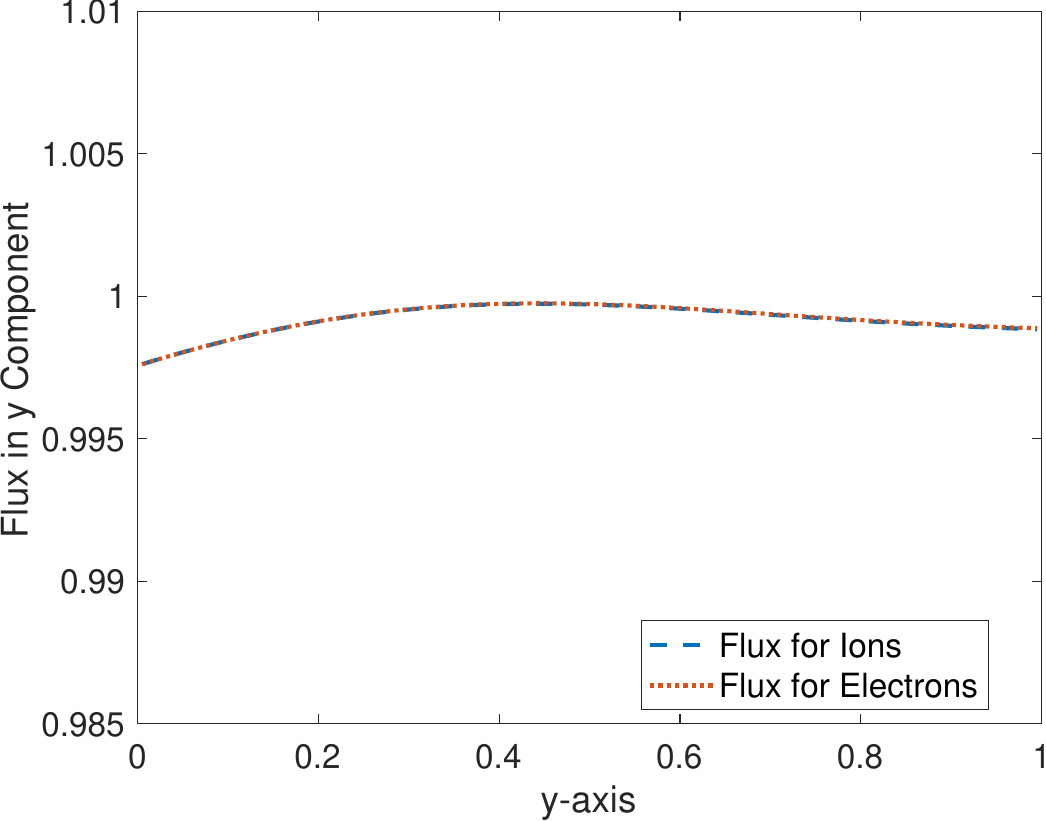}}%
  \hspace*{0.04\textwidth}
  \subfigure[AP: $(n_\alpha \mathbf{u}_\alpha)_{y}$, $t=10$]{\includegraphics[width= 0.3\textwidth]{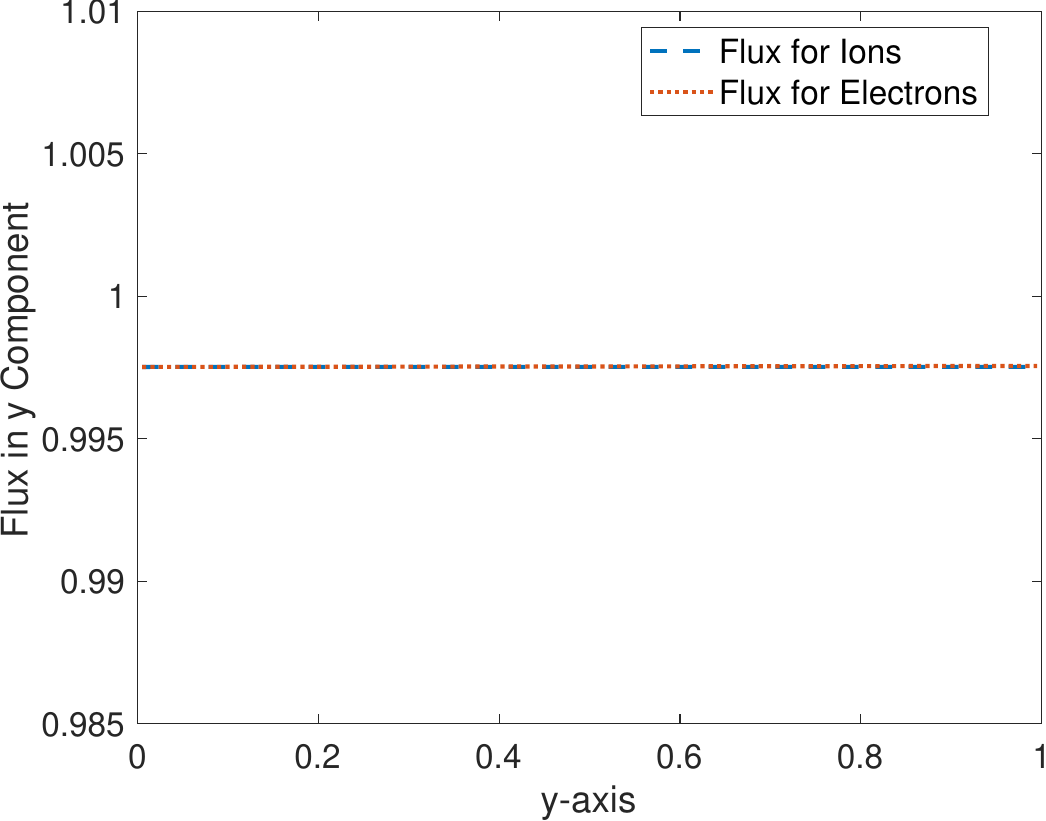}}
  \end{center}
  \caption{Perturbed 1D initial data: Slices (at {$x=0$}) of the ion and electron densities ($n_e,\ n_i$), electric potential ($\phi$) and  y-component of the ion and electron momentum ($(n_\alpha \mathbf{u}_\alpha)_{y}$) as functions of $y$. These quantities are computed by the AP-scheme at different (dimensionless) times.\label{fig:test_perturbation_density_phi} }
  \end{figure}

The density, electric potential as well as the $y$-component of the momentum obtained thanks to the AP scheme using the perturbed initial densities are plotted on Fig.~\ref{fig:test_perturbation_density_phi}. 
  The pressure gradient along the magnetic field lines creates an acceleration of the electrons and ions, while the electric field adjusts to preserve the quasi-neutrality of the plasma. The perturbation is gradually smeared out (see Figs.~\ref{fig:test_perturbation_density_phi}(a-c)) and a steady state is recovered: the perturbation amplitude is reduced by a factor 5 at $t=1$, the perturbation being vanished at time t=10, as observed on Fig.~\ref{fig:test_perturbation_density_phi}(c),(f),(i).

  To conclude this section, let us emphasize the necessity of using the AP scheme. According to Proposition~\ref{prop:unbounded:parallel:momentum} (see Sec.~\ref{sec:num:meth}), the residual issued from the numerical resolution of  the implicit scheme is amplified by the squared electron Mach number reciprocal ($\varepsilon_{M,e}^{-1}=M_e^{-2}$), which in turn affects the parallel momentum. In the example above, the direct linear solver MUMPS \cite{doi:10.1137/S0895479899358194} is used for the numerical resolution of the linear systems. This solver achieves a relative residual of $10^{-10}$, close to the computer arithmetic precision, for the small system sizes considered herein. Given that the electron Mach number is equal to $10^{-3}$ ($\varepsilon_{M,e}^{-1}=10^6$), the amplified relative residual is roughly $10^{-4}$. This already indicates a loss of precision. However, in this simple benchmark run on coarse meshes, the numerical results of the implicit scheme remain acceptable and consistent with those obtained using the AP-scheme.

  However, the merits of the different schemes must be assessed in the more general framework of three-dimensional applications on refined meshes. In this context, direct solvers are far less efficient, and iterative methods (preconditioned Krylov methods) are generally preferred. It then becomes much more challenging to achieve residuals as low as those obtained here, highlighting the relevance of the AP scheme. 
 In contrast to the implicit scheme and according to Proposition~\ref{prop:bounded:parallel:momentum}, the AP scheme is free from any residual amplification, thereby ensuring an accurate parallel momentum approximation.

\begin{figure}[htbp]
\begin{center}
\includegraphics[width=0.5\textwidth]{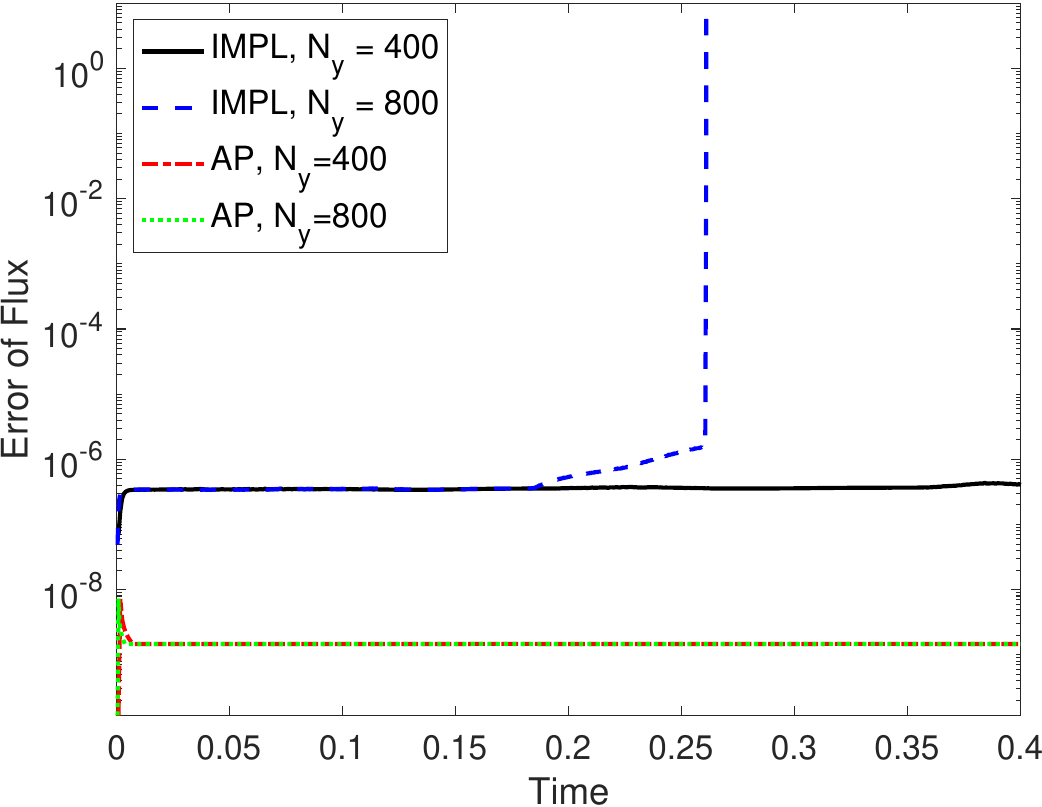} 
\end{center}
\caption{Un-perturbed 1D initial data: Error history (infinity norm) on the electron parallel momentum carried out by either the implicit or the AP scheme for the mesh size ($N_y$ being the number of cells) and $\varepsilon_{M,e}=10^{-6}$. The computations are performed using a time step controlled by a 0.32 CFL condition ($\Delta t = 0.32 \Delta y$).\label{fig:perturbationMe}}
\end{figure}

 This issue can be examined in more detail using Fig.~\ref{fig:perturbationMe}, which shows the time evolution of the error in the parallel momentum computed by both the implicit and AP schemes for various mesh resolutions.
The approximation provided by the AP scheme maintains an accuracy that is almost unaffected by mesh resolution and remains stable over time. In contrast, the error in the approximation computed by the implicit scheme is significantly amplified and, for refined meshes, blow up after only a few time steps. This conclusively demonstrates the benefit of using a scheme that prevents the amplification of numerical errors by the asymptotic parameter.

\subsection{Collision with neutral particles}
The HIT-PSI experiment is designed to study the interaction of a hot plasma with neutral particles. The model is thus supplemented with the minimal upgrades to account for these interactions, the details of which are specified in \appendixname~\ref{sec:Collisions}. 
The boundary conditions as well as the initial particle density, velocities and temperatures are resumed from Sec.~\ref{sec:num_low_Mach} (without perturbations). 

Two frequencies are introduced. First, the recombination frequency of particles of specie $\alpha$ with neutrals, denoted $\hat \nu_\alpha$. Second, the momentum exchange frequency of specie $\alpha$ with neutrals at rest, denoted $\nu_\alpha$ (see \appendixname~\ref{sec:Collisions}). For both the recombination and collision rates, the following function of the aligned coordinate $y$ is considered 
\begin{align*}
\nu(y) = \frac{1}{2} (1 + \tanh(5-10y)).
\end{align*}
This choice mimic the distribution of neutral particles injected near the target, the recombination frequency and the collision frequency are therefore strong near $y=1$ and weaker near the source ($y=0$), the function $\nu$ being close to 1 for $y=1$ while vanishing in $y=0$. Two different settings are investigated
\begin{itemize}
\item Setting 1: $\hat\nu_i = 0.6 \nu$, $\hat\nu_e= 0.1 \nu$, and $\nu_i=\nu_e=\nu$;
\item Setting 2: $\hat\nu_i = 0.7 \nu$, $\hat\nu_e= 0.1 \nu$, and $\nu_i=\nu_e=\nu$.
\end{itemize}
Note that, the recombination frequency is a decreasing function of the particle temperature \cite{2006CoPP...46....3S}, hence in any of the settings $\hat\nu_i >\hat\nu_e$.

\begin{figure}[htbp]
\begin{center}

\subfigure[$t=5$]{\includegraphics[width= 0.3\textwidth]{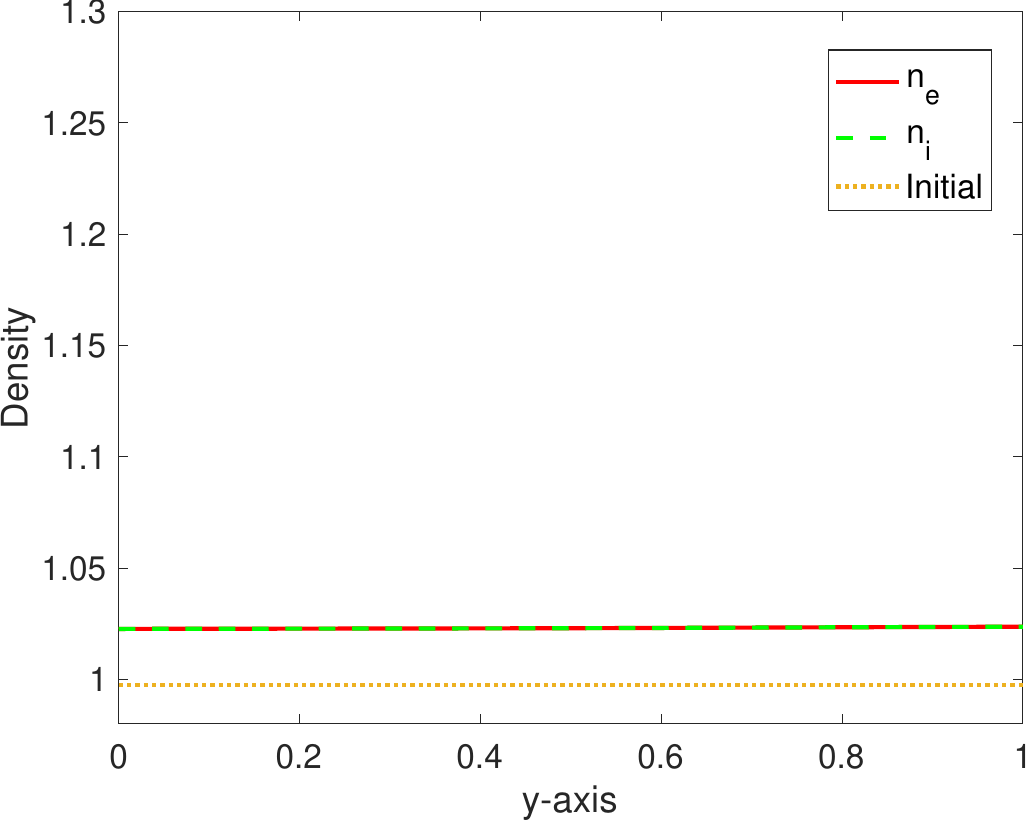}}
\subfigure[$t=7.5$]{\includegraphics[width= 0.3\textwidth]{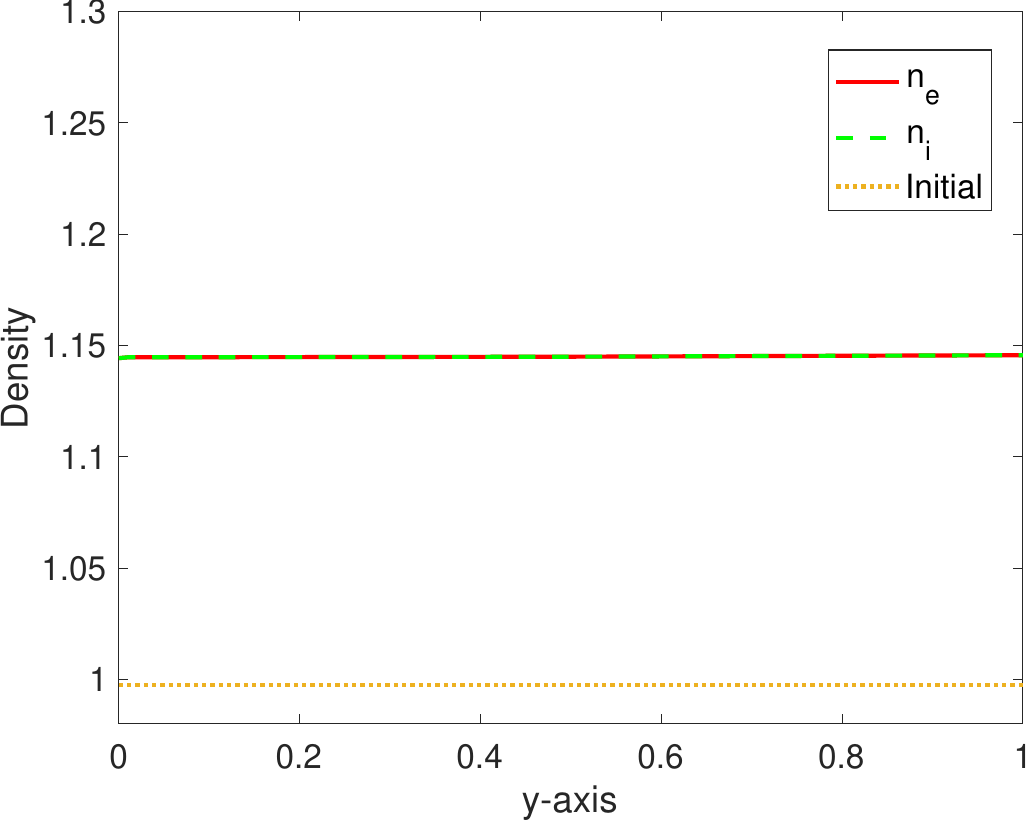}}
\subfigure[$t=10$]{\includegraphics[width= 0.3\textwidth]{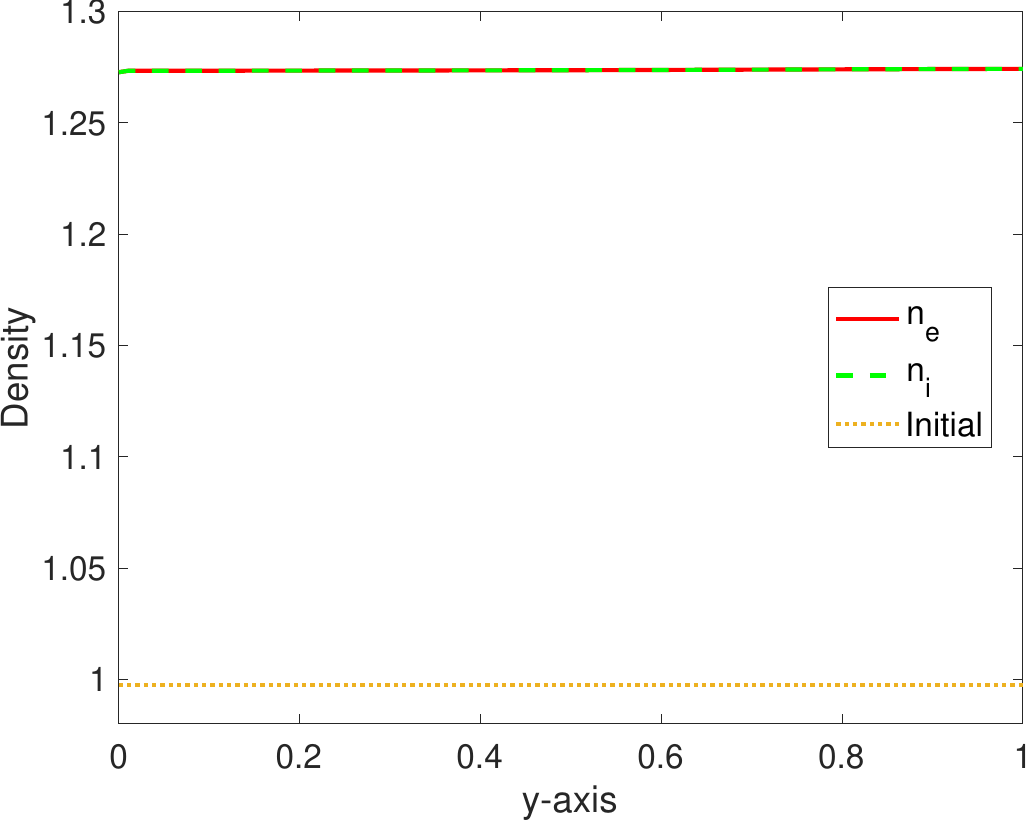}}

\subfigure[$t=5$]{\includegraphics[width= 0.3\textwidth]{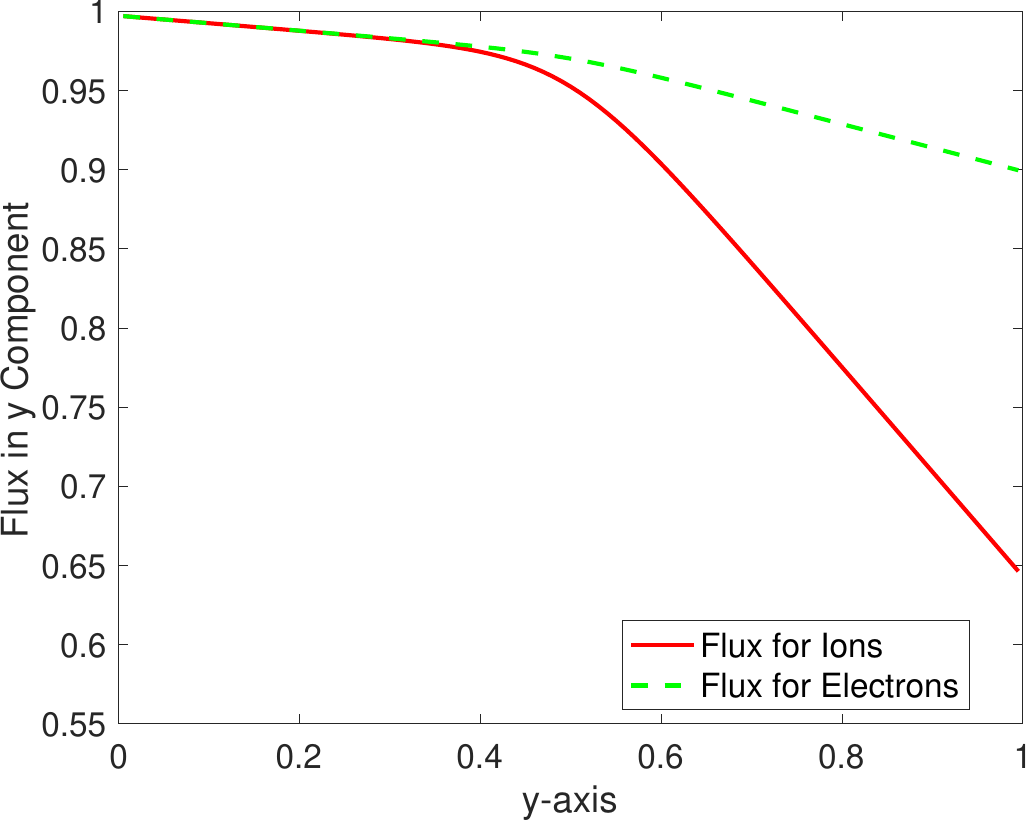}}
\subfigure[$t=7.5$]{\includegraphics[width= 0.3\textwidth]{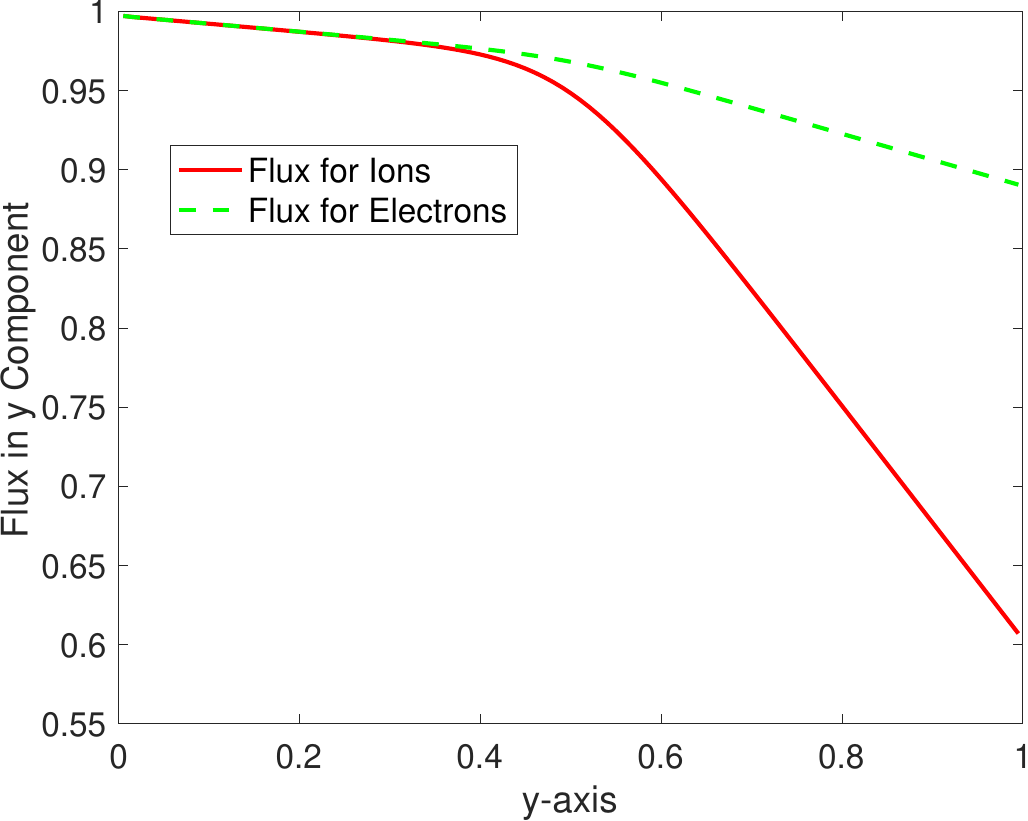}}
\subfigure[$t=10$]{\includegraphics[width= 0.3\textwidth]{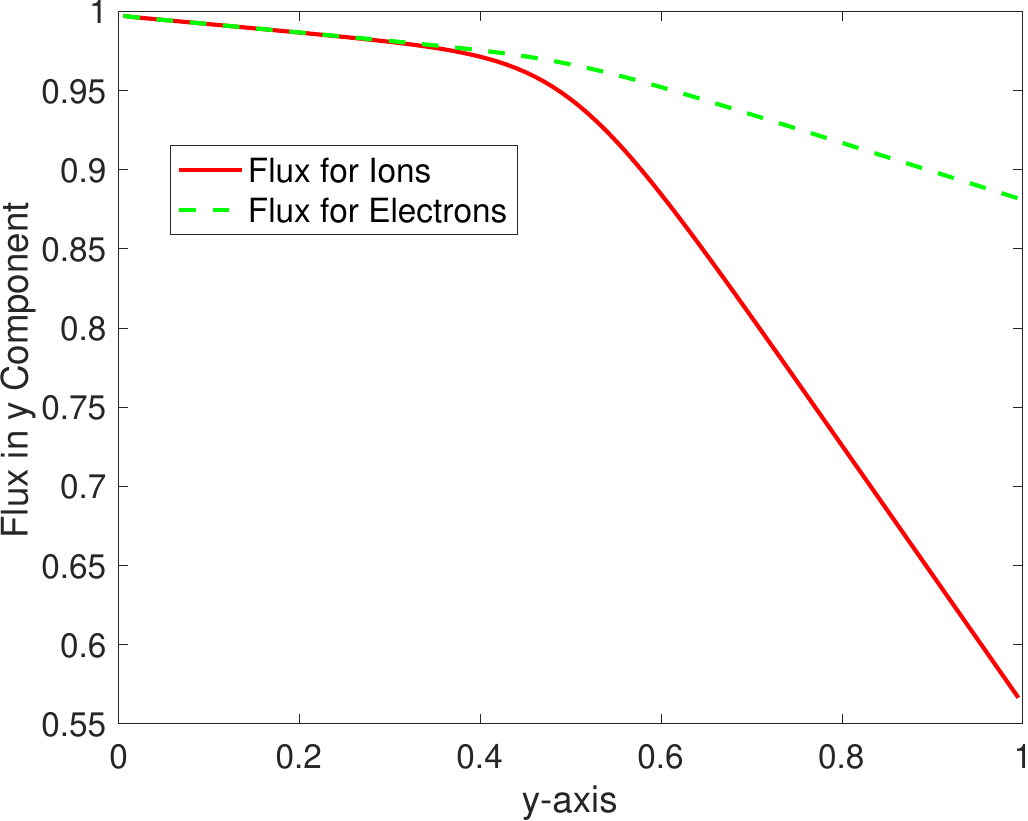}}

\end{center}
\caption{Interaction with neutrals - Setting~1: Slices of the density and aligned momentum $(n_\alpha \mathbf{u}_\alpha)_y$  at $x=0$ for different (dimensionless) times. \label{fig:test_collision_slice:One}}
\end{figure}

The simulations are carried out on a mesh with $N_x\times N_y = 20\times 100$ cells, the time step being $\Delta t = 10^{-3}$. 
The density and parallel momentum related to setting~1 are displayed on Fig.~\ref{fig:test_collision_slice:One}. The parallel momentum decreases (see Figs.~\ref{fig:test_collision_slice:One} (d)-(f)) over time, due to collisions with neutrals at rest. This prevents the plasma from being effectively expelled at the target, leading to plasma accumulation.
This entails an increase of the electron and ion densities over time as depicted by Figs.~\ref{fig:test_collision_slice:One} (a)-(c)). The electron and ion parallel momentum is also observed to be inconsistent in region of frequent collisions ($y>0.5$). This is explained by the lower electron recombination rate in comparison to ions. To maintain quasi-neutrality in the plasma, the ion momentum near the target is therefore lower than the electron momentum.


\begin{figure}[htbp]
  \begin{center}
  
  \subfigure[$t=5$]{\includegraphics[width= 0.3\textwidth]{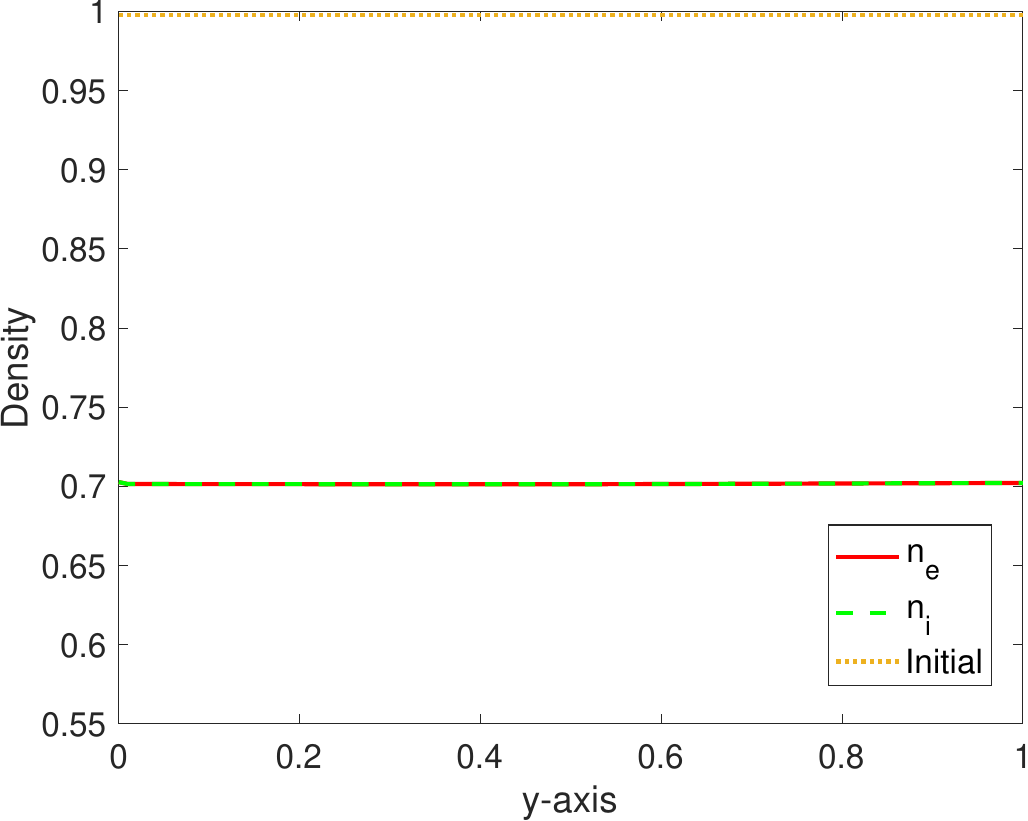}}
  \subfigure[$t=7.5$]{\includegraphics[width= 0.3\textwidth]{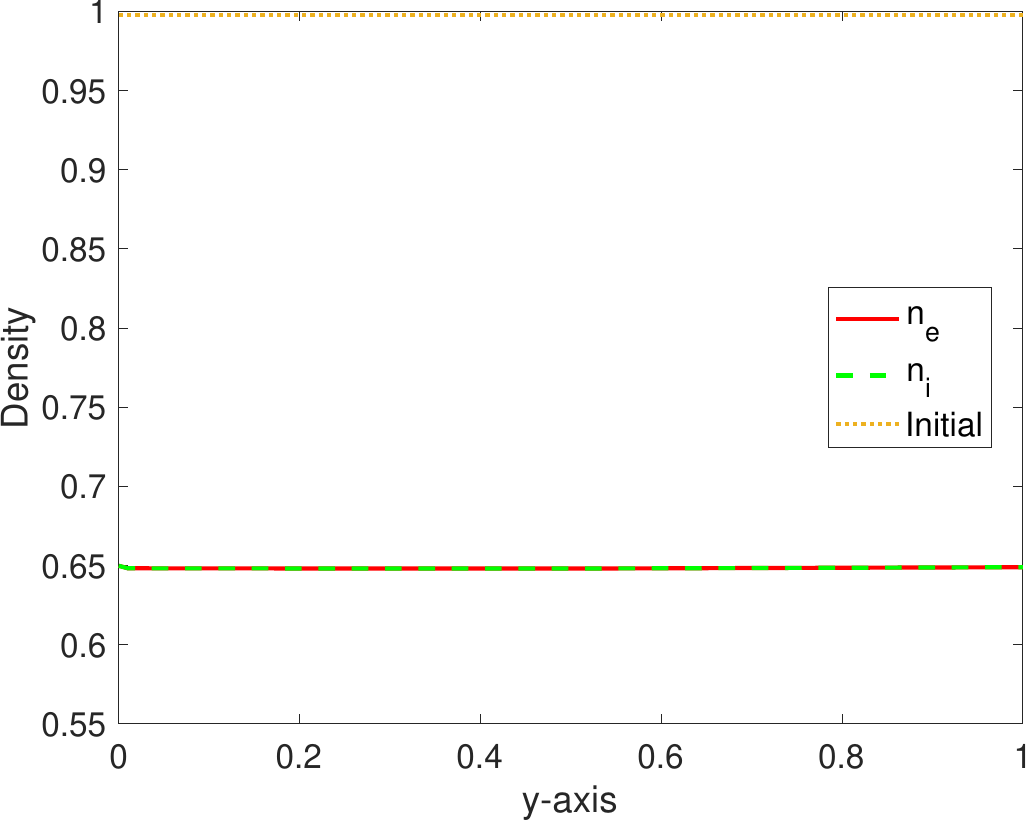}}
  \subfigure[$t=10$]{\includegraphics[width= 0.3\textwidth]{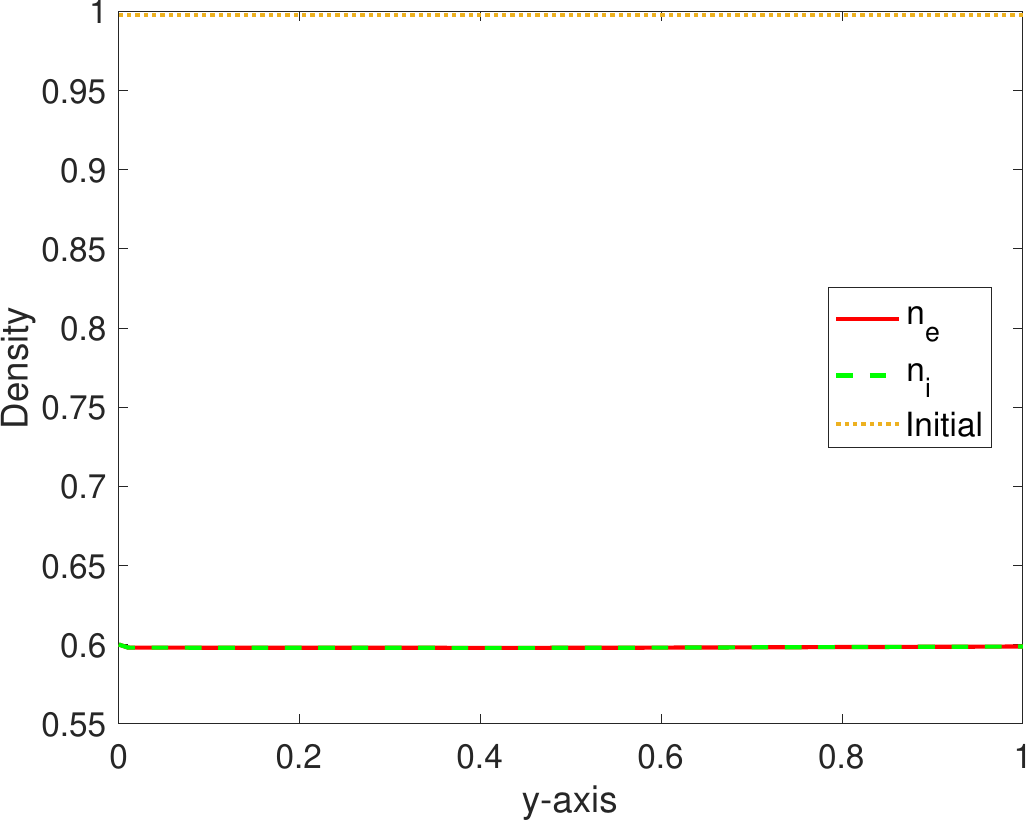}}

  \subfigure[$t=5$]{\includegraphics[width= 0.3\textwidth]{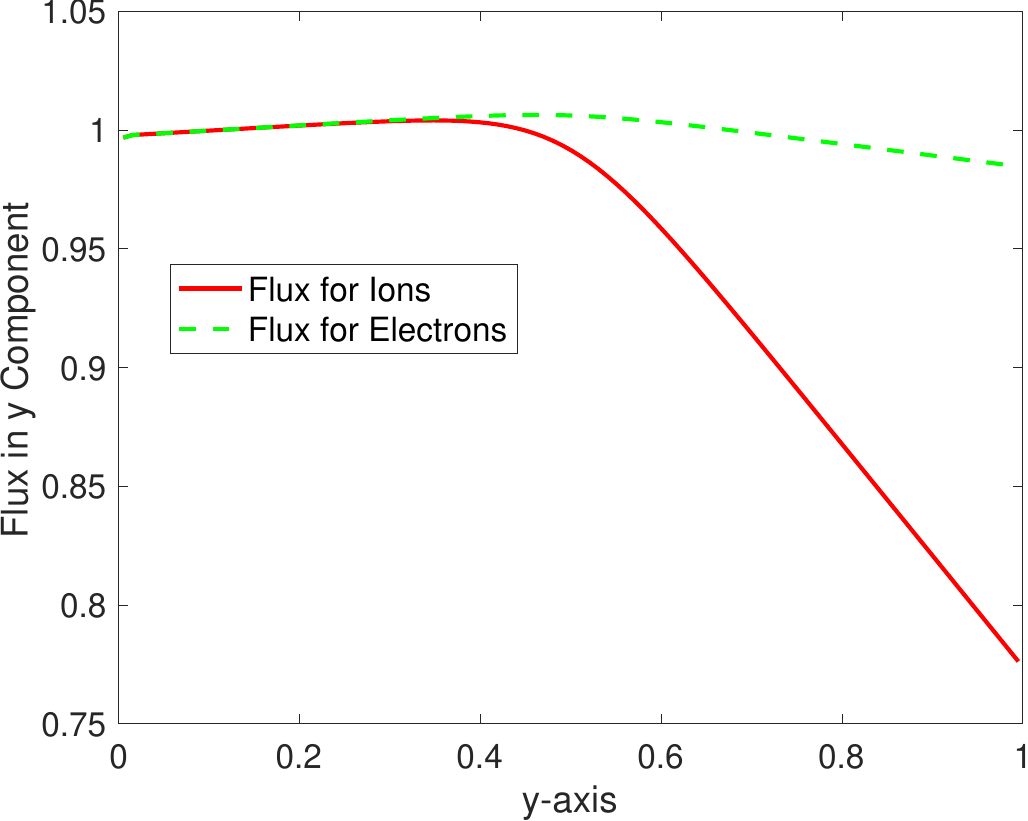}}
  \subfigure[$t=7.5$]{\includegraphics[width= 0.3\textwidth]{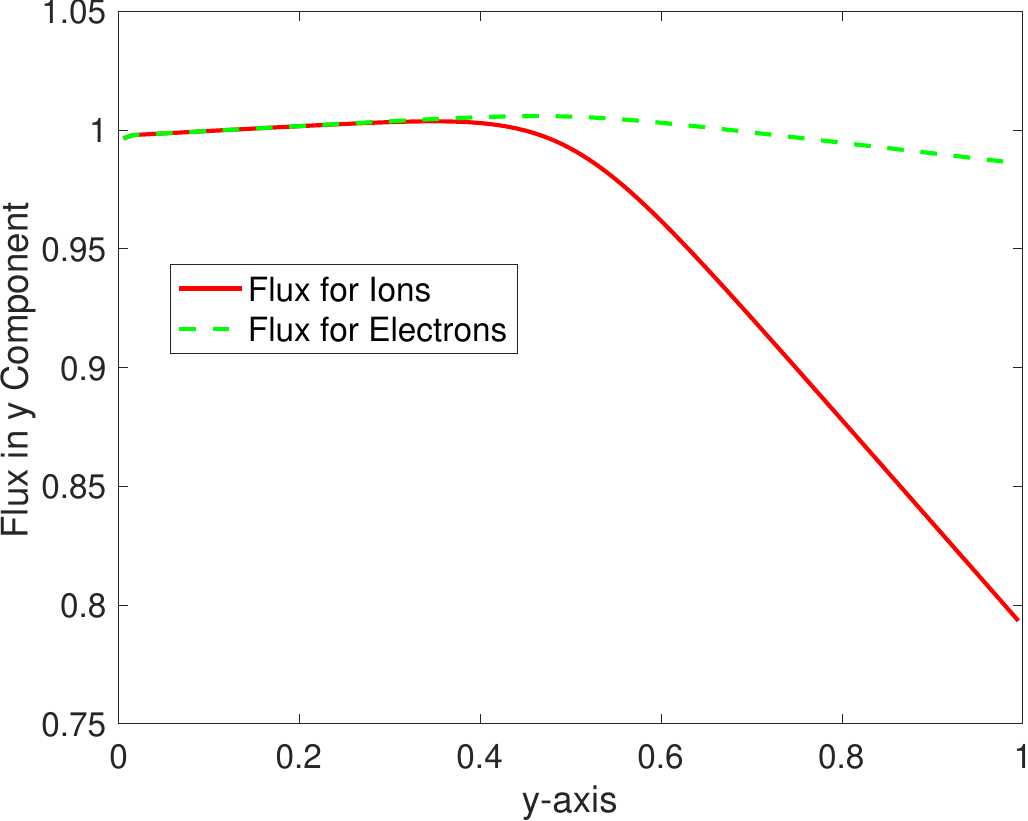}}
  \subfigure[$t=10$]{\includegraphics[width= 0.3\textwidth]{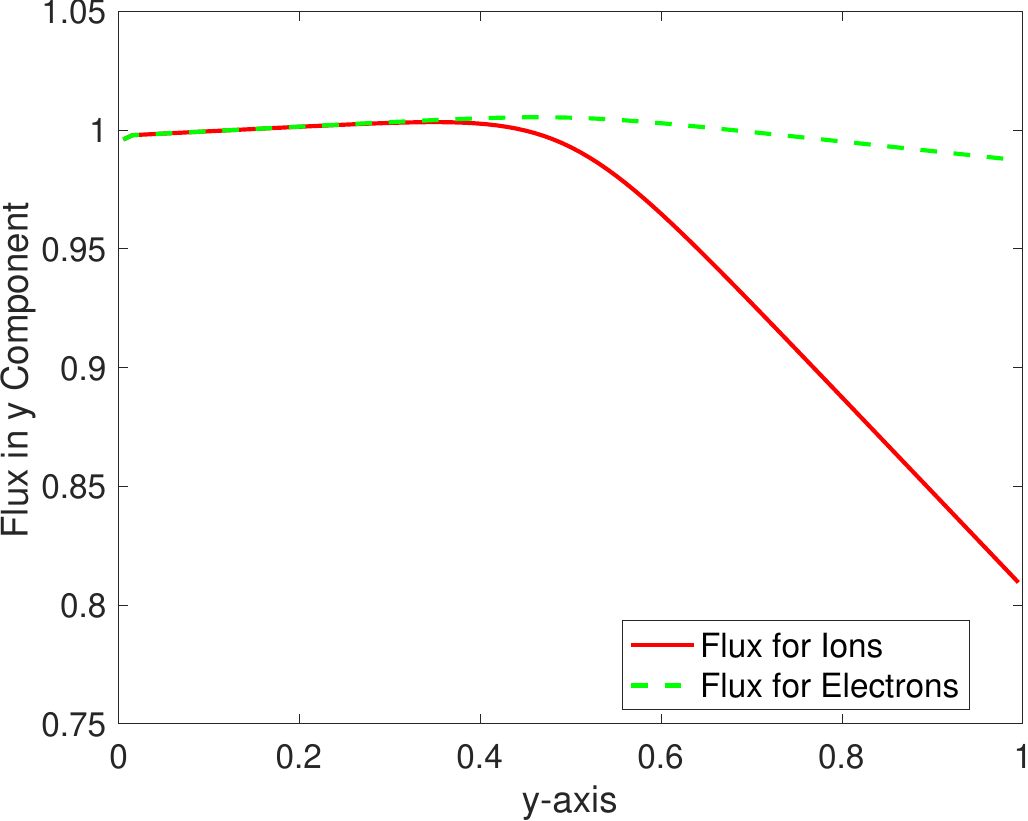}}

  \end{center}
  \caption{Interaction with neutrals - Setting~2: Slices of the density and aligned momentum $(n_\alpha \mathbf{u}_\alpha)_y$ at $x=0$ for different (dimensionless) times.\label{fig:test_collision_slice:Two}}
  \end{figure}

For setting~2, the recombination rates are increased entailing a diminishing of the particle  density over time (see Fig.~\ref{fig:test_collision_slice:Two}(a)-(c)). The parallel momentum component is a non-monotone function of the aligned coordinate, first increasing then decreasing (see Figures~\ref{fig:test_collision_slice:Two} (d)-(f)). The ion flux is decreasing with time due to the drop of plasma density. In contrast, the electron parallel flux at the target is observed to remain roughly constant over time.

From these two cases, we observe that due to the presence of neutral particles, plasma momentum exhibits a significant decreasing trend near the target, which is consistent with numerical simulation expectations.

  \subsection{Collision-less isothermal sheath}
  This test case is aimed at assessing the ability of the AP scheme to cope with local quasi-neutrality breakdowns, by comparing against the theoretical results of~\cite{Riemann_2005} and the  numerical experiments conducted in \cite{alvarez-laguna_asymptotic_2020}. For this, a quasi one-dimensional framework is considered, consisting of quantities homogeneous along the $x$ and $z$ directions. Following ~\cite{alvarez-laguna_asymptotic_2020}, the  continuity equations for the ions and electrons is supplemented with an ionization source term yielding
  \begin{align*}
  \partial_t n_\alpha + \nabla\cdot (\mathbf{u}_\alpha n_\alpha) = n_e \nu^{iz},\quad \alpha\in\{e,i\}\,,
  %
  \end{align*}
  where the ionization frequency is defined by
  \begin{align*}
  \nu^{iz}(x) := \frac{|q_{i,y}(y=0)| + |q_{i,y}(y=1)|}{\int_0^1 n_e dy}.
  \end{align*}

\begin{figure}[htbp]
    \begin{center}
    \subfigure[Density $(n_\alpha)$]{\includegraphics[width=0.47\textwidth]{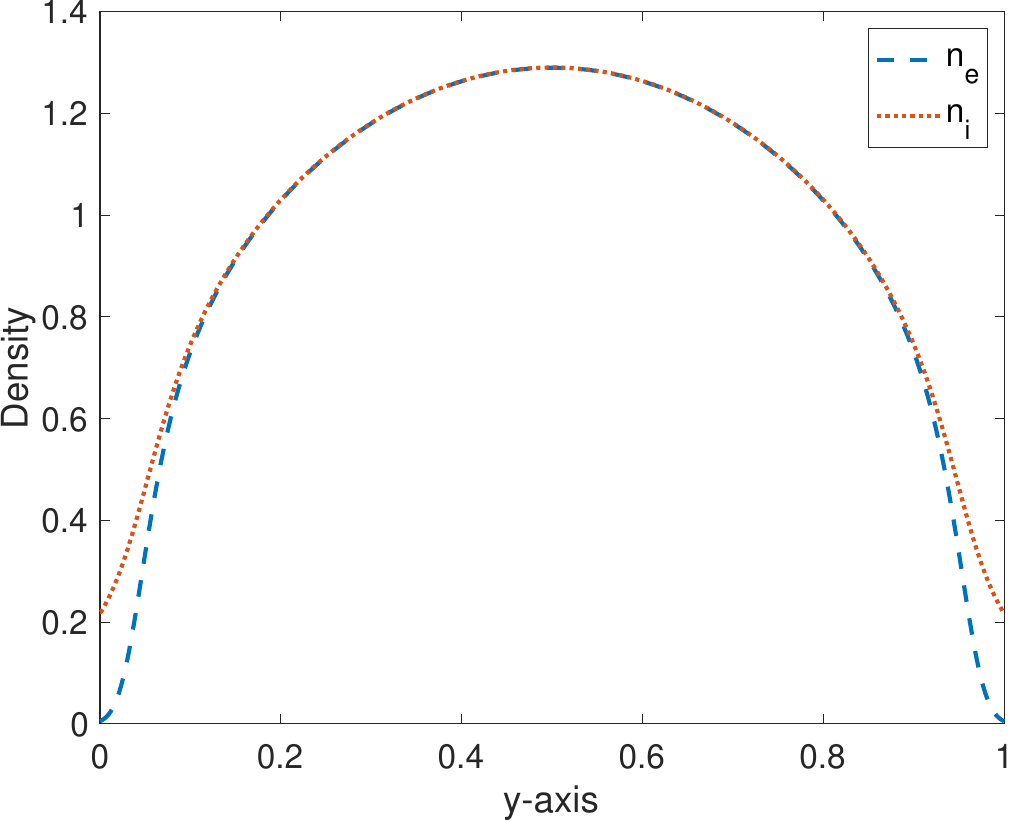}}%
    \hspace*{0.06\textwidth}%
    \subfigure[Momentum $(n_\alpha\mathbf{u}_\alpha)_y$]{\includegraphics[width=0.47\textwidth]{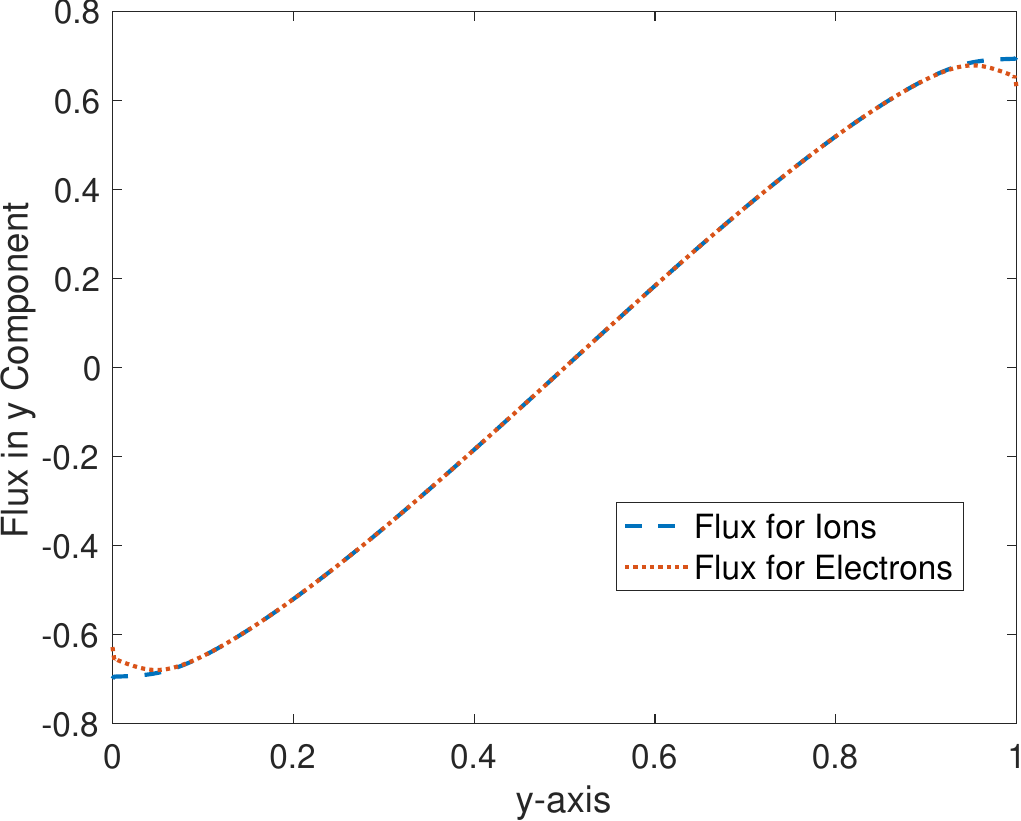}}
    
    \subfigure[Velocity $(\mathbf{u}_\alpha)_y$]{\includegraphics[width= 0.47\textwidth]{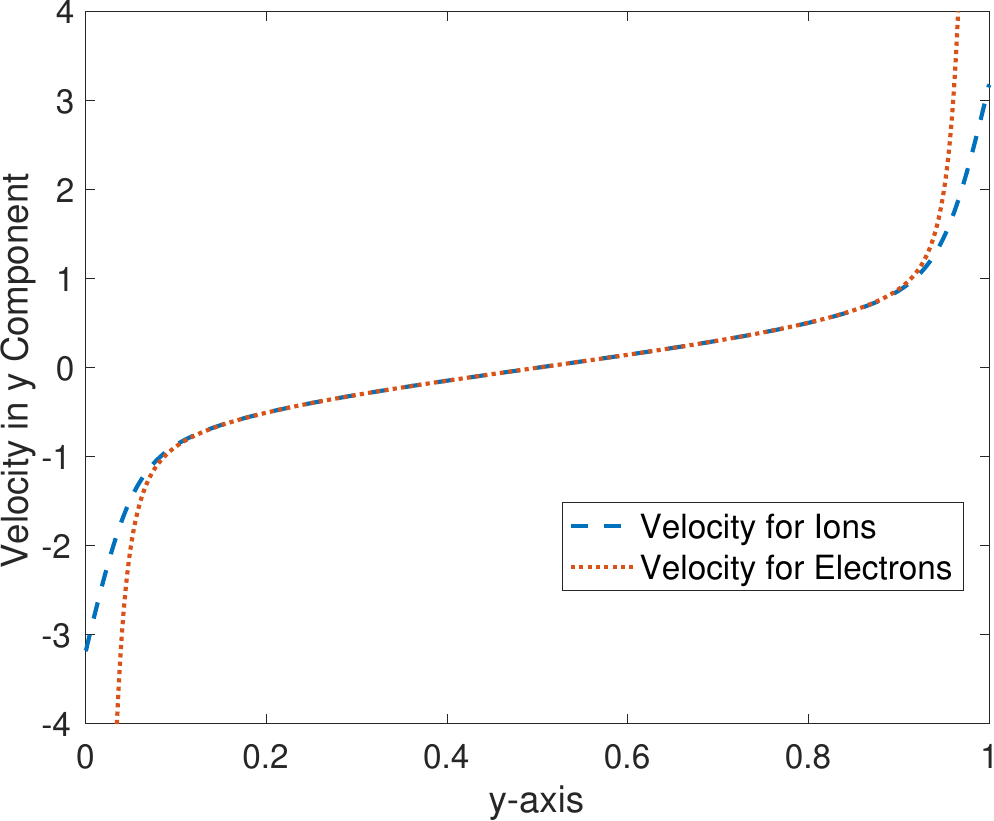}}%
    \hspace*{0.06\textwidth}\subfigure[Shifted electric potential ($\phi-\|\phi\|_\infty$)]{\includegraphics[width= 0.47\textwidth]{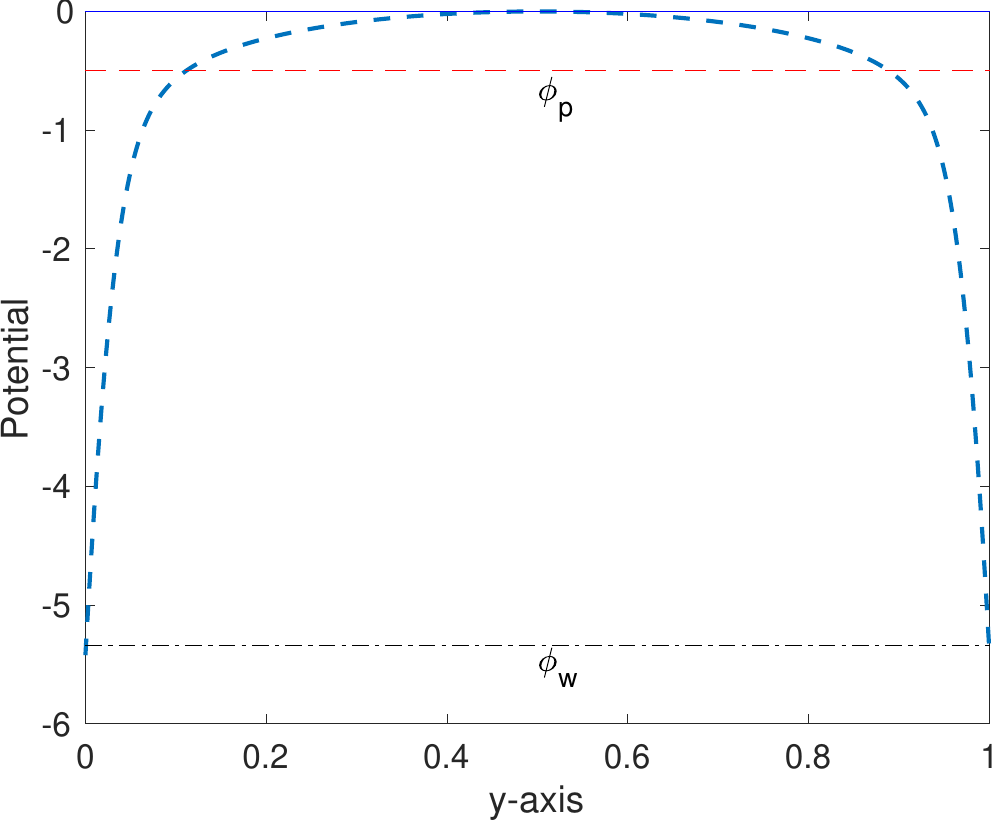}}
    \end{center}
    \caption{Collision-less isothermal sheath: Plasma and electric potential characteristics at the steady state (t=5) for computations carried out by the AP scheme with $1000$ cells and a time step $\Delta t =5\times10^{-6}.$ \label{fig:test_sheath_1Dtest}}
    \end{figure}

  The boundary conditions are axis-symmetric at $x=0$; at $x=0.2$ Neumann boundary conditions are used for $n_\alpha,\phi,q_{\alpha,x},q_{\alpha,z}$. At $y=0$ or $y=1$, Dirichlet boundary conditions are prescribed for $n_e, \phi, \kappa_\alpha, q_{\alpha,y}$, while Neumann boundary conditions are used for $n_i$. The flux of electrons  at the boundaries ($y=0$ or $y=1$) are prescribed according to \cite{alvarez-laguna_asymptotic_2020}
  \begin{align*}
  n_e\mathbf{u}_e(y=0,t) = - \frac{n_e}{\sqrt{2\pi\varepsilon}},\quad \,n_e\mathbf{u}_e(y=1,t) =\frac{n_e}{\sqrt{2\pi\varepsilon}} \,,
  \end{align*}
  where the parameter $\varepsilon$, representing the electron to ion mass ratio, is set to $10^{-5}$.
  The scaled Debye $\lambda$ length value is $10^{-2}$ obtained for a uniform plasma density equal to 1 (in dimensionless unit). The quasi-neutrality may be broken with a local decrease of the plasma density. The main scales characterizing this test case are 
  \begin{align*}
  \bar{x} = 3\times10^{-2} \text{m},\quad \bar{t} = 1.6\times10^{-5} \text{s},\, \quad \bar{T_e} = 1 \text{eV}\,,\quad \bar{T_i} = 0.025 \text{eV}\,.
  \end{align*}
  The plasma is initially  at rest with velocities equal to $(0,0,0)$. The mesh is composed of  $N_x\times N_y = 3\times 1000$ cells, the time step  $\Delta t = 5\times 10^{-6}$, defining a mean CFL number, during the simulation, close to $0.4$.  The numerical results performed with the AP scheme are presented on Fig.~\ref{fig:test_sheath_1Dtest} at $t=5$, at this time a steady state being reached.
  
  It is noticed that all the quantities maintain perfect homogeneity along the $x$-direction preserving the one-dimensional evolution of the plasma in accordance with the analytic framework proposed in \cite{Riemann_2005}. The evolution of the plasma is governed by a non-quasi-neutral physics. Indeed, the electron density departs significantly from the ion one near the two electrodes located at ($y=0$ and $y=1$) as displayed by the plots of Fig.~\ref{fig:test_sheath_1Dtest}(a). The local charge density creates an electric field, as depicted by the plot of the electric potential on Fig.~\ref{fig:test_sheath_1Dtest}(d). The plot of this figure displays the shifted potential $\phi-\|\phi\|_\infty$ together with the pre-sheath potential drop $\phi_\text{p}$ and the wall potential drop $\phi_\text{w}$ \cite{Riemann_2005, alvarez-laguna_asymptotic_2020} defined as 
  \begin{align*}
  \phi_\text{p} = \frac{k_B \bar{T_e}}{2e},\quad \phi_\text{w} = \frac{k_B \bar{T_e}}{2e}\ln\left( \frac{1}{2\pi \varepsilon}\right).
  \end{align*}
  The particle fluxes are plotted on Fig.~\ref{fig:test_sheath_1Dtest}(b). The ion and electron fluxes match, except in the vicinity of the electrodes. The amplitude and extent of this discrepancy are very similar to the results obtained in \cite{alvarez-laguna_asymptotic_2020} with the same mesh resolution. It vanishes with the increase in mesh resolution. 
  
  Overall, the outputs of the AP-scheme are in a very good agreement with both the theoretical estimates and the numerical experiments conducted in \cite{alvarez-laguna_asymptotic_2020} and demonstrate the ability to account precisely for local quasi-neutrality breakdowns.

\section{Conclusion and perspectives}\label{Conclusions}
In this paper, a simplified mathematical model is proposed to investigate the physics of magnetically confined plasmas. The model is based on an isothermal fluid description of the plasma coupled with an electrostatic equation, under the assumption that the external magnetic field remains constant in time. The typical scales of the targeted application underscore the multiscale nature of the problem—where the plasma is assumed to be quasi-neutral and the electrons are described by the drift approximation. However, these assumptions may not hold in localized regions, thereby motivating the development of specialized numerical methods within the class of asymptotic-preserving schemes.

The asymptotic-preserving property is carefully analyzed for two distinct discretizations. Selected terms are discretized implicitly to overcome the degeneracy of the equations in the asymptotic regimes, leading to the solution of linear systems and the introduction of errors associated with limited computer arithmetic, as reflected by the residuals from these linear system numerical resolutions.

The novelty of this study lies in incorporating residual errors into the evaluation of the asymptotic-preserving properties of various discretizations. In particular, the analysis demonstrates that these residuals are amplified by the asymptotic parameter associated with the drift regime. Although this amplification is marginal for one- or two-dimensional problems on small meshes when using direct solvers, it becomes significant for three-dimensional problems on refined meshes, where iterative solvers—yielding higher residuals—are typically employed. Furthermore, the analysis reveals that the asymptotic-preserving property can be compromised by a nonconforming discretization of second-order differential operators, particularly in the context of collocated spatial discretization. This observation calls for the development of AP schemes that are robust against the amplification of approximation errors, including those stemming from computer arithmetic, such as the one proposed herein.

Numerical experiments demonstrate the merits of the proposed method across various test cases. The results highlight its capability to handle local quasi-neutrality breakdowns, deliver a parallel momentum approximation whose accuracy is independent of the asymptotic parameters, and partially reproduce the physical phenomena observed in the HIT-PSI experiment, which defines the target application.

To further advance this work, the model must be extended to incorporate at least an energy equation to relax the isothermal assumption inherent in the current minimal model, as well as enhancements in the numerical efficiency of the proposed AP scheme to support three-dimensional computations using iterative solvers equipped with dedicated block preconditioners.

%
  \appendix
\section{Rescaled quantities}\label{sec:RQ}
The linear plasma experiment HIT-PSI \cite{wang2023} provides the typical application targeted within this work. The characteristic length and time are that of the device and of its operating, the scale of the physical quantities are reported into Tab.~\ref{tab:physics}. 
%
%
%
%
\begin{table}[htbp]
  \begin{center}
  \caption{Main characteristics of the targeted application operations~\cite{wang2023}.\label{tab:physics}}
  \begin{tabular}{|c|rc|}\hline
$\bar{x}$ & 1 &m \\
$\bar{t}$ & $10^{-3}$ &s  \\
$\bar{u}$ & $10^3$ &$\text{m}\cdot\text{s}^{-1}$ \\
$\bar{T}=\bar{T}_\alpha$ & $10^4$ &K\\
$\bar{B}$ & 2 &T\\
$\bar{E}$ & $10^6$ &$\text{V}\cdot \text{m}^{-1}$ \\
$\bar{n}$ & $10^{20}$ &$\text{m}^{-3}$ \\ \hline
  \end{tabular}
  \begin{tabular}{|l|rc|}\hline
    $\lambda_D$ & $7 \cdot 10^{-7}$ &(m) \\
    $C_{s,e}$ & $4 \cdot 10^5$ &($\text{m}\cdot\text{s}^{-1}$)\\
    $C_{s,i}$ &  $ 10^2$ & ($\text{m}\cdot\text{s}^{-1}$)\\
  $\omega_{c,e}$ &  $4\cdot 10^{11}$ & ($\text{s}^{-1}$) \\
  $\omega_{c,i}$ & $2\cdot 10^{8}$ & ($\text{s}^{-1}$) \\ \hline
  \end{tabular} 
  \begin{tabular}{|c|c|}\hline
    $\lambda$ & $10^{-6}$\\
    $M_e$ & $10^{-3}$\\
    $M_i$ & $10^{-1}$\\
    $\Omega_{e}$ & $10^{8}$ \\
$\Omega_{i}$ & $10^{5}$  \\
$\eta$ & $10^6$\\\hline
  \end{tabular}
\end{center}
\end{table} 
\section{Collisions with neutral particles}\label{sec:Collisions}
The model is supplemented with specific source terms to account for collisions with neutrals. The continuity equation related to the specie $\alpha$ reads
\begin{equation*}
n_\alpha^{k+1} = n_\alpha^k - \Delta t \nabla\cdot(n_\alpha \mathbf{u}_\alpha)^{k+1} - \Delta t\hat\nu_\alpha n^{k+1}_\alpha,\label{eq:discret_ni_collision}\\
%
\end{equation*}
where $\hat\nu_\alpha$ is the (isothermal) recombination frequency with neutral particles for either the ions ($\alpha=i$) or the electrons ($\alpha=e$). Similarly, the momentum equations  are supplemented with forces accounting for the exchange of momentum with neutrals assumed to be at rest, yielding
\begin{align}
&\begin{multlined}[0.9\textwidth]
  (n_\alpha\mathbf{u}_\alpha)^{k+1} = (n_\alpha\mathbf{u}_\alpha)^{k} - \Delta t\nabla\cdot(n_\alpha \mathbf{u}_\alpha   \otimes\mathbf{u}_\alpha)^k  \\[0.3em]
 - \frac{\Delta t}{\varepsilon_{M_\alpha}} \left(\nabla P_\alpha^{k+1} +  \bar q_\alpha \eta n_\alpha^k E^{k+1} \right)+ \bar q_\alpha\frac{\Delta t}{\varepsilon_{\Omega_\alpha}} (n_\alpha\mathbf{u}_\alpha)^{k+1}\times B -  \Delta t\nu_\alpha (n_\alpha u_\alpha)^{k+1},\end{multlined}\label{eq:discret_nui_collision}
%
\end{align}
where $\bar q_i =+1 $, $\bar q_e =-1 $ and $\nu_\alpha$ is the collision frequency against neutral for particles of specie $\alpha$. The collisions introduce a modification of the mobility matrices, with
\begin{align}
  &\begin{multlined}[0.9\textwidth]
(n_\alpha\mathbf{u}_\alpha)^{k+1} =\mathbb{A}_\alpha \left[  (\widetilde{n_\alpha\mathbf{u}_\alpha})^{k} - \frac{\Delta t}{\varepsilon_{M_\alpha}} \nabla P_\alpha^{k+1} + \bar q_\alpha \frac{\Delta t}{\varepsilon_{M_\alpha}}\eta n_\alpha^k E^{k+1} \right],\label{eq:discret_nui2_collision}
%
  \end{multlined}\\
 &(\widetilde{n_\alpha\mathbf{u}_\alpha})^{k}  = (n_\alpha\mathbf{u}_\alpha)^{k} - \Delta t\nabla\cdot(n_\alpha \mathbf{u}_\alpha   \otimes\mathbf{u}_\alpha)^k  \,,
\end{align}
where
\begin{align*}
\mathbb{A}_i &= 
\begin{pmatrix}
\frac{h_i\varepsilon_{\Omega_i}^2}{h_i^2\varepsilon_{\Omega_i}^2+\Delta t^2} & 0 & -\frac{\Delta t \varepsilon_{\Omega_i}}{h_i^2\varepsilon_{\Omega_i}^2+\Delta t^2}\\
0 & \frac{1}{h_i} & 0\\
\frac{\Delta t \varepsilon_{\Omega_i}}{h_i^2\varepsilon_{\Omega_i}^2+\Delta t^2} & 0 & \frac{h_i\varepsilon_{\Omega_i}^2}{h_i^2\varepsilon_{\Omega_i}^2+\Delta t^2}
\end{pmatrix},
\mathbb{A}_e = 
\begin{pmatrix}
\frac{h_e\varepsilon_{\Omega_e}^2}{h_e^2\varepsilon_{\Omega_e}^2+\Delta t^2} & 0 & \frac{\Delta t \varepsilon_{\Omega_e}}{h_e^2\varepsilon_{\Omega_e}^2+\Delta t^2}\\
0 &  \frac{1}{h_e} & 0\\
-\frac{\Delta t \varepsilon_{\Omega_e}}{h_e^2\varepsilon_{\Omega_e}^2+\Delta t^2} & 0 & \frac{h_e\varepsilon_{\Omega_e}^2}{h_e^2\varepsilon_{\Omega_e}^2+\Delta t^2}
\end{pmatrix},
\end{align*}
and
\begin{align*}
h_i = 1+ \Delta t \nu_i, \quad h_e = 1+ \Delta t \nu_e.
\end{align*}

Simple algebra yields the following definition for the parallel momentum
\begin{eqnarray*}
(n_\alpha \mathbf{u}_{\alpha})_{y}^{k+1} = \frac{1}{1 + \nu_\alpha\Delta t} (\widetilde{n_\alpha \mathbf{u}_{\alpha}})_y^{k} -  \frac{\Delta t}{\varepsilon_{M_\alpha} (1+ \nu_\alpha\Delta t)}(T_\alpha \partial_y n^{k+1}_i - \bar q_\alpha\eta n_\alpha^k E_y^{k+1})\,.
%
\end{eqnarray*}
Collisions with neutral tend to slow down the particles is the parallel direction a property well accounted for this last expression of the parallel momentum.
\bibliographystyle{abbrv}
\bibliography{bib}
\end{document}